\documentclass[aps,llncs,preprint]{revtex4-2}
\usepackage{graphicx}
\usepackage{amsmath}
\usepackage{float}
\begin{document}
	
\title{Spectroscopic evidence of multi-gap superconductivity in non-centrosymmetric AuBe} 

\author{Soumya Datta$^1$, Aastha Vasdev$^1$, Partha Sarathi Rana$^2$, Kapil Motla$^3$, Anshu Kataria$^3$, Ravi Prakash Singh$^3$, Tanmoy Das$^2$, and Goutam Sheet$^1$}

\email{goutam@iisermohali.ac.in}

\affiliation{$^1$Department of Physical Sciences, Indian Institute of Science Education and Research Mohali, Sector 81, S. A. S. Nagar, Manauli, PO 140306, India}

\affiliation{$^2$Department of Physics, Indian Institute of Science, Bangalore 560012, India}

\affiliation{$^3$Department of Physics, Indian Institute of Science Education and Research Bhopal, Bhopal 462066, India}

\begin{abstract}
	
AuBe is a chiral, non-centrosymmetric superconductor with transition temperature $T_C$ $\simeq$ 3.25 K. The broken inversion symmetry in its crystal structure makes AuBe a possible candidate to host a mixed singlet-triplet pairing symmetry in its superconducting order parameter ($\Delta$). This possibility was investigated by transport, thermodynamic, and muon-spin rotation/relaxation experiments in AuBe. However, this issue was not addressed using direct spectroscopic probes so far. In addition, certain ambiguities exist in the description of superconductivity in AuBe based on $\mu$SR experiments reported earlier. Here we report scanning tunneling spectroscopy (STS) on AuBe down to 300 mK. We found a signature of two superconducting gaps (with 2$\Delta_1/k_{B}T_{C}$ = 4.37 and 2$\Delta_2/k_{B}T_{C}$ = 2.46 respectively) and a clean BCS-like temperature dependence of both the gaps. We have also performed band structure calculations to identify the different bands that might give rise to the observed two-gap superconductivity in AuBe.

\end{abstract}

\maketitle


The nature of superconductivity for a non-centrosymmetric superconductor (NCSC) varies widely from material to material. Evidence of line nodes in the superconducting gap is found in numerous NCSCs like Mo$_3$Al$_2$C\cite{bauer2010unconventional}, CeIrSi$_3$\cite{mukuda2008enhancement}, CePt$_3$Si\cite{bonalde2005evidence,frigeri2004superconductivity},   Li$_2$Pt$_3$B\cite{yuan2006s,nishiyama2007spin}, and recently Ru$_{7}$B$_{3}$\cite{datta2021spectroscopic}. On the other hand, in a number of  NCSCs, only a conventional, isotropic, s-wave, fully-gapped superconducting phase is reported. Nb$_{0.18}$Re$_{0.82}$\cite{karki2011physical}, BiPd\cite{sun2015dirac,yan2016nodeless},  T$_2$Ga$_9$ (T = Rh, Ir)\cite{shibayama2007superconductivity,wakui2009thermodynamic} and LaMSi$_3$ (M = Rh\cite{anand2011specific}, Ir\cite{anand2014physical}, Pd, Pt\cite{smidman2014investigations}) are a few examples of that. There are also NCSCs where the superconducting phase is not well described either by a fully open gap or a nodal gap opened in a single band. In such a situation, a multi-gap model becomes necessary. LaNiC$_2$\cite{chen2013evidence}, TaRh$_2$B$_2$\cite{mayoh2018multigap} and Ln$_2$C$_3$ (Ln = La,Y)\cite{sugawara2007anomalous,kuroiwa2008multigap} are examples of such NCSCs. In this context, AuBe with transition temperature $T_C$ $\sim$ 3.25 K\cite{matthias1959superconductivity,rebar2015exploring,amon2018noncentrosymmetric,singh2019type,beare2019mu,rebar2019fermi} is interesting due to multiple reasons. The material contains the heavy element Au and the crystal structure of the material is non-centrosymmetric (cubic space-group symmetry of $P2_13$)\cite{rebar2015exploring,amon2018noncentrosymmetric}. This makes AuBe a potential candidate for a mixed pairing state\cite{frigeri2004superconductivity,yuan2006s,datta2021spectroscopic}. Further, the crystal structure of AuBe is also chiral ($B20$, FeSi type), and such structure is predicted to host chiral fermions\cite{bradlyn2016beyond,tang2017multiple,chang2017unconventional}. To note, the $B20$ structure is also the only known crystal structure where bulk magnetic skyrmions are reported\cite{yu2011near,tonomura2012real,kanazawa2012possible}. Further, according to a recent study, the superconducting pairing of AuBe originates from multiple energy bands\cite{khasanov2020multiple,khasanov2020single}. All these collectively make AuBe an attractive system where a search for unconventional superconducting pairing and exotic quasiparticle excitations are warranted.\\
 
There are two primary outstanding issues within and beyond the previous reports on superconducting AuBe. The first one is about the type of superconductivity with respect to its magnetic properties. Amon $et$ $al.$\cite{amon2018noncentrosymmetric} reported type-II (with Ginzburg-Landau parameter $k_{GL}$ = 0.4) superconductivity while Singh $et$ $al.$\cite{singh2019type} and Beare $et$ $al.$\cite{beare2019mu} reported type-I ($k_{GL}$ = 2.3) behavior in AuBe based on their independent dc magnetization, specific heat, and muon-spin rotation/relaxation ($\mu$SR) studies. Reber $et$ $al.$\cite{rebar2015exploring,rebar2019fermi} attempted to resolve the issue revealing a crossover from type-II to type-I superconductivity\cite{tinkham2004introduction,annett2004superconductivity} at $\sim$ 1.2 K based on their resistivity, dc magnetization, ac susceptibility and specific heat studies. This suggests the possibility that AuBe falls under the class of type II/I superconductors, the behavior which was for a long time investigated in intentionally disordered elemental superconductors with $k_{GL}$ $\sim$ 1/$\sqrt{2}$\cite{krageloh1969flux,auer1973magnetic}. Recently, such crossover was also proposed in superconductors with non-centrosymmetric crystal structures and/or multiple bands participating in superconductivity\cite{babaev2017type,samoilenka2020spiral}. Coincidentally, both these properties happen to be relevant in the present context of AuBe. The second issue lies in the description of the superconducting order parameter. All the reports up to 2019, including those which have mutually conflicting conclusions about the type (type I vs. type II) of superconductivity in AuBe, have agreement on the fact that AuBe is a pure, isotropic s-wave, spin-singlet superconductor\cite{amon2018noncentrosymmetric,singh2019type,beare2019mu,rebar2019fermi}. However, Khasanov $et$ $al.$\cite{khasanov2020multiple,khasanov2020single}, in their two successive papers in 2020 proposed an entirely new unconventional multi-gap mechanism in AuBe. Based on new $\mu$SR experiments\cite{khasanov2020multiple}, the authors reported that the temperature evolution of the thermodynamic critical field $B_C$ of AuBe could not be explained without considering at least two different gaps. From a self-consistent two-gap model, it was found that the corresponding 2$\Delta/k_{B}T_{C}$ were 4.52 and 2.37, respectively. Further, based on a detailed comparative analysis between the usual single gap model and a self-consistent two-gap model, the latter’s superiority over the former was concluded\cite{khasanov2020single}. Hence, in order to acquire a decisive evidence of multi-band superconductivity or to rule out such a possibility in AuBe, it remains an important task to measure the superconducting gap(s), spectroscopically in this system. In addition, detailed studies of the temperature and magnetic field evolution of the superconducting order parameter are also warranted to probe the nature of the gaps.\\


The polycrystalline sample of AuBe was prepared by arc melting method from a stoichiometric mixture of elemental Au and Be. The preparation and characterization details are published elsewhere\cite{singh2019type}. To directly probe the superconducting gap(s) in AuBe, we employed low-temperature scanning tunneling microscopy (STM) and spectroscopy (STS) in a \textit{Unisoku} system with \textit{RHK R9} controller, inside an ultra-high vacuum (UHV) cryostat kept at $\sim$ $10^{-10}$ mbar. The lowest temperature down to which the measurements were performed was 310 mK. As such experiments are extremely sensitive to the surface cleanliness, a few layers from the surface were first removed by mild reverse sputtering in an argon environment \textit{in-situ}, inside another UHV preparation chamber that is connected to the main STM/S chamber. Then the sample with its pristine surface was transferred to the scanning stage at low temperature for experiments. The Tungsten (W) tip, which was prepared outside by electrochemical etching, was also cleaned by high-energy electron-beam bombardment inside the same UHV preparation chamber. This whole process helped us probe the pristine surface of AuBe. Spectra were further recorded at random points on the surface, and at each point on the sample surface, we found a clean spectrum with a fully opened gap. The differential conductance $dI/dV$ was measured using a lock-in based ac modulation technique (amplitude 40 $\mu$V, frequency 3 kHz).\\



All the spectra we probed show two clear peaks symmetric about $V$ = 0. The position of these coherence peaks provides a direct measure of the superconducting energy gap ($\Delta$). However, depending on the different points on the sample surface probed, the positions of the coherence peaks vary from $\pm$250 $\mu$V to $\pm$375 $\mu$V, approximately. In Figure 1, we show eight representative tunneling spectra captured at different points on the surface of AuBe, all at the lowest temperature. The spectra presented in Figure 1(a)-(d) visually have a wider spectral gap compared to the spectra presented in Figure 1(e)-(h). Intrinsic disorder on the surface of the sample can exhibit such variation, and that was indeed our primary guess. Nevertheless, to extract the exact value of the order parameter $\Delta$ for each spectrum, we proceeded further with the analysis. All the spectra were first normalized $w.r.t$ conductance at 1.5 mV, where they are almost flat. Then these experimental spectra were compared with numerically generated spectra using the expression for tunneling current within a single-gap model given by the following equation.\\

$I(V) \propto \int_{-\infty}^{+\infty}N_s(E)N_n(E-eV)[f(E)-f(E-eV)]dE$\\

Here, $N_s(E)$ and $N_n(E)$ are respectively the normalized density of states (DOS) of the superconducting sample and the normal metallic tip where $f(E)$ is the Fermi-Dirac distribution function\cite{bardeen1957theory}. Within this single band model, $N_s$ is given by the following expression of the Dynes formula\cite{dynes1978direct}.\\

$N_s(E) \propto Re\left(\frac{(E-i\Gamma)}{\sqrt{(E-i\Gamma)^2-\Delta^2}}\right)$\\

In our analysis, $\Gamma$ takes care of all possible reasons for spectral broadening, including that due to finite quasiparticle lifetime and other possible interband and intraband scattering effects. Though this model is routinely used and highly successful to analyze the tunneling spectroscopic data of a conventional superconductor, it fails to match any experimental spectrum of AuBe, irrespective of whether the spectrum was visibly wider (like in Figure 1(a)-(d)) or narrower(like in Figure 1(e)-(h)). To explain the issue, two such theoretical plots (red and blue curves) are shown on each experimental spectrum (black circles) presented in Figure 1(d) and (h). The red line represents the closest match for the upper portions of each spectrum, especially at the coherence peaks. However, that fails to match the lower portion near $V$ = 0, underestimating the actual depth of the spectrum. Since $\Gamma$ is fixed arbitrarily without a complete knowledge about the microscopic origin of the same, we first tried to fit the spectra by brute force making $\Gamma$ free. In order to match the lower part, if we tune the parameter $\Gamma$ down and adjust the $\Delta$ as needed too, we face a situation (blue line) where the lower part of the theoretical plot matches perfectly with the experimental spectrum, but now deviates significantly above and overestimates the actual height of the coherence peaks. Hence, it becomes clear that though a quasiparticle excitation spectrum of AuBe looks like a standard one with a pair of clear coherence peaks, it cannot be explained within the framework of a single-gap $s$-wave model. Replacing the isotropic $s$-wave $\Delta$ with an anisotropic $\Delta Cos(n\theta)$\cite{tanaka1995theory} (where the integer $n$ can be 1, 2 or higher depending on $p$, $d$ or higher-order symmetries, respectively) does not give an acceptable description of the data. Instead, doing that gives rise to a sharp ‘V’-shaped spectrum (not shown in the figure) in contrast to the experimentally obtained ‘U’-shaped ones. Also, we did not find a gapless flat spectrum or a spectrum with a zero-bias conductance peak anywhere on the surface. Hence we considered the simplest two-gap model for our spectra which was further motivated by the facts about AuBe reported in the past. Following is a description of such facts.\\


First, based on the de Haas–van Alphen experiments performed on AuBe, the presence of multiple bands crossing the Fermi level was reported by Rebar $et$ $al.$\cite{rebar2015exploring,rebar2019fermi}. Second, based on density functional theory (DFT) and band-structure calculations performed independently by Rebar $et$ $al.$\cite{rebar2015exploring,rebar2019fermi} and Amon $et$ $al.$\cite{amon2018noncentrosymmetric} such band crossing was confirmed for at least three conductive bands. However, though this special type of band structure creates a possibility of multiband superconductivity in AuBe, it can not be taken as an evidence of the same. We also noted the reports by Khasanov $et$ $al.$\cite{khasanov2020multiple,khasanov2020single}, where the authors extracted the thermodynamic critical field $B_C$ from their $\mu$SR experiment on AuBe and explained its temperature dependence based on a self-consistent two-gap model. In such a scenario, the quasiparticle excitation spectrum for a two-band superconductor can be determined simply by adding the two single gap BCS spectra for the two respective bands\cite{suhl1959bardeen}. Considering such a picture, the quasiparticle DOS of the $j$-th band can be written as follows.\\

 $N_{s,j} (E) = N_j(E_F) Re\left(\frac{(E-i\Gamma_j)}{\sqrt{(E-i\Gamma_j)^2-\Delta_{0j}^2}}\right)$, $j$ = 1, 2\\

Here $j$ is the band index, $N_j(E_F)$ is the normal state DOS at the Fermi level corresponding to the $j$th band, and $\Delta_{0j}$ is the amplitude of the superconducting energy gap formed in the $j$th band. The tunneling current, which has contributios from both the bands, will now take the following form.\\

 $I(V) \propto \sum_{j = 1,2}\alpha_j\int_{-\infty}^{+\infty}N_{sj}(E)N_n(E-eV)[f(E)-f(E-eV)]dE$\\
 
 Here $\alpha_j$ is the relative contribution of the $j$-th band to the tunneling current. When we tried to fit the spectra using this model, it became extremely successful over the entire energy range. We have presented such theoretical plots with green lines in Figure 1(d) and (h). It is clear that such a plot matches both the upper and lower part of each spectrum this time very well. The extracted values of two superconducting gaps ($\Delta_1$ and $\Delta_2$) and the corresponding two broadening parameters ($\Gamma_1$ and $\Gamma_2$) are also mentioned for each spectrum. To note, the pair of the superconducting gap values did not vary noticeably from $\Delta_1$ $\simeq$ 320$\pm$10 $\mu$eV, and $\Delta_2$ $\simeq$ 180$\pm$10 $\mu$eV across the spectra. The parameters which actually do vary are the relative contributions of each band to the total tunneling current, i.e $\alpha_1$ and $\alpha_2$. For the spectra represented in Figure 1(a)-(d), we found  $\alpha_1 > \alpha_2$, and for the spectra represented in Figure 1(e)-(h), the opposite.\\

In MgB$_2$, the two gaps are distinctly visible from experiments like Andreev reflection spectroscopy, tunneling spectroscopy, etc.\cite{szabo2001evidence,giubileo2001two,schmidt2002evidence,iavarone2002two,gonnelli2002direct,silva2015tunneling}. In such experiments, two pairs of coherence peaks corresponding to the two gaps appear in the quasiparticle DOS. In contrast, the spectra that we recorded on AuBe do not have distinct multi-gap features, and visually they look like a single-gap BCS spectrum. However, from our detailed analysis, we found that a usual single band model cannot explain such spectra while a simple two-gap model can. Visually the spectra of AuBe appear different from that of MgB$_2$, primarily, because the amplitudes of the two gaps in AuBe are close. Schopohl $et$ $al.$\cite{schopohl1977tunneling} and Noat $et$ $al.$\cite{noat2010signatures} explained such situations with interband scattering and tunneling of quasiparticles. They described multiple characteristic features of the multi-band spectra like damped quasiparticle peaks, kinks near the peaks, dips beyond the peaks, etc. that may appear as a consequence of such interband physics. From a close visual inspection of our spectra, when we compare the experimental data (black circles) with the best single gap fits (blue lines) in Figure 1(d) and (h), we actually can notice the first two features mentioned above. However, it is important to note that a simple multi-gap model proposed by Suhl $et$ $al.$\cite{suhl1959bardeen} is successful to fit (green lines) our experimental data with very high fidelity, and in this model, more complicated factors like interband scattering, k-selective tunneling etc. were not taken into account. At the same time, when Iavarone \textit{et al.}\cite{iavarone2002two} reported a distinct two-gap superconductivity in MgB$_2$ by tunneling spectroscopy, Eskildsen \textit{et al.}\cite{eskildsen2002vortex} also reported tunneling spectroscopy on the same material. Interestingly, when the latter group attempted to explain their spectra using the usual single gap Dynes equation, their theoretical fit overestimated the coherence peak. This can be compared with our fittings (blue lines) in Figure 1(d) and (h) using the single gap Dynes model. Now, we also note that the existence of two gaps cannot be proved simply based on the analysis of certain spectra. If two different bands participate in the superconductivity of AuBe, they are expected to evolve with temperature and external magnetic field independently and differently, unless the inter-band scattering is too strong.\\


We present the temperature ($T$) dependence of a typical spectrum in Figure 2(a), where the colored circles represent the experimentally obtained spectra. With increasing $T$, the coherence peaks gradually decrease, and all the gap features disappear at 1.77 K. The corresponding theoretical fits within the two-band model\cite{suhl1959bardeen} are shown on top of each experimental spectrum as black lines where the values of $\alpha_1$(0.8) and $\alpha_2$(0.2) were kept unchanged over the entire $T$ range. The two gaps extracted from the fits are plotted with $T$ in Figure 2(b) with red ($\Delta_1$) and blue ($\Delta_2$) triangles. Each gap independently follows a BCS-like dependence\cite{bardeen1957theory} up to 1.77 K, where they merge and disappear. The smaller gap ($\Delta_2$) slightly deviates from the BCS-line at higher $T$, which is typical in the presence of a non-zero inter-band scattering\cite{suhl1959bardeen}. Such a situation, where two gaps independently evolve with temperature until they disappear at the same $T_C$, is valid for a multi-gap superconductor where interband scattering is negligible. This further validates the model we have used for our analysis. The extracted values of the two broadening parameters ($\Gamma_1$ and $\Gamma_2$) are presented in the inset of Figure 2(b) throughout the $T$ range. To note, the $\Gamma$s do not increase with temperature as expected for a strong coupling superconductor. Rather they decrease slightly with increasing $T$ or remain almost constant considering the same energy scale of $\Delta$s. In the pioneering paper by Dynes $\textit{et al.}$\cite{dynes1978direct}, the authors pointed out that such behavior is a describing feature of a weakly coupled superconductor (like Al) vis-a-vis a strongly coupled one (like PbBi). Our observation is consistent with the previous reports\cite{amon2018noncentrosymmetric,rebar2019fermi,singh2019type} that AuBe is a weakly coupled superconductor with a specific heat jump near $T_C$, $\Delta C/\gamma_nT_C \approx$ 1.26, and coupling strength $\lambda_{e-p} \approx$ 0.5. From our two-gap analysis of the tunneling spectra, we found that 2$\Delta_1/k_{B}T_{C}$ = 4.37 and 2$\Delta_2/k_{B}T_{C}$ = 2.46, respectively. The former one is slightly underestimated, while the latter one is slightly overestimated compared to the values 4.52 and 2.37 respectively, as was reported by Khasanov $\textit{et al.}$ based on $\mu$SR experiments\cite{khasanov2020multiple}. In the presence of a very small yet finite interband scattering, these differences are in accordance with the two-gap model\cite{suhl1959bardeen}.\\


To gain further information about the pairing mechanism, we now focus on the magnetic field dependence of the spectra. The colored circles in Figure 3(a) show the experimentally obtained spectra, all measured at 310 mK, and the black lines represent the corresponding two-band fits. All the superconducting features, including the coherence peaks, disappear at 17 kG. The evolution of the extracted two gaps (larger $\Delta_1$ and smaller $\Delta_2$) with magnetic field are shown in Figure 3(b). Up to 10 kG, both the gaps tend to decrease slowly in a linear fashion, and beyond that, they decrease faster until becoming zero at 17 kG. Two important conclusions can be drawn from this observation. First, the gradual transition supports the type-II behavior in AuBe as reported by Amon $et$ $al.$\cite{amon2018noncentrosymmetric} but contradicts the type-I behavior reported by Singh $et$ $al.$\cite{singh2019type} and Beare $et$ $al.$\cite{beare2019mu}. However, this contradiction can be easily resolved considering the proposed type II/I superconductivity by Reber $et$ $al.$\cite{rebar2019fermi}. As our magnetic measurements are performed at $\sim$ 310 mK, which is far below the crossover point 1.2 K\cite{rebar2015exploring,rebar2019fermi}, the type-II behavior is normal and visible. On the second note, the reported critical field $H_C$ of AuBe from various bulk measurements varies between 259 G to 335 G\cite{amon2018noncentrosymmetric,rebar2019fermi,singh2019type,beare2019mu,khasanov2020multiple}, but from our magnetic field dependence, we found the local critical field $H_{C(l)}$ as high as 17 kG. It is interesting to note that Reber $et$ $al.$\cite{rebar2019fermi} also reported a considerably higher resistive upper critical field $H_{C(\rho)}$ compared to the thermodynamic upper critical field $H_C$ found from heat capacity and magnetization measurements. To explain this enhancement, the authors have eliminated the possibility of defects or impurity phases at the surface and concluded it to be an intrinsic surface behavior with a possible topological protection. However, based on our data, such a possibility can be neither confirmed nor ruled out. The field dependence of $\Gamma_1$ and $\Gamma_2$ are presented in the inset of Figure 3(b). With increasing magnetic field, $\Gamma_1$ increases almost linearly where, $\Gamma_2$ increases slowly up to $\sim$10 kG and then increases faster.\\


In order to understand the origin of multiband superconductivity in AuBe, we investigated the band structure, density of states, and the Fermi surface of the system through first-principles electronic structure calculations. Ab-initio electronic structure of AuBe was calculated using the Density Fuctional Theory (DFT)\cite{kohn1965self} implemented in Quantum Espresso\cite{giannozzi2009quantum}. The calculated band structure and the topology of the Fermi surfaces are consistent with the previous reports\cite{amon2018noncentrosymmetric,rebar2019fermi}. In our calculations, the experimental crystal structure was used with a relaxed cell parameter of 4.709$\AA$. A full-relativistic pseudo-potential with Perdew-Burke-Ernzerof (PBE)\cite{perdew1996generalized} exchange-correlation potential was used in the projected augmented wave(PAW)\cite{blochl1994projector} method, both with and without spin-orbit coupling (SOC). Self-consistent charge-density convergence was achieved on a $10 \times10 \times10$ Monkhorst-Pack\cite{monkhorst1976special} $k$-grid. The energy cut-off for the calculation is $60$ $Ry$ and the Fermi-surfaces and Femi-velocities reported here were obtained on a $k$-grid of $20\times20\times20$ without the SOC.\\

In Figure 4(a), we have plotted the Fermi surface of AuBe, where the corresponding Fermi velocity is shown with color gradients. The band dispersion along the high symmetry directions and the orbital projected DOS are plotted in Figure 4(c) and 4(d), respectively. The corresponding high symmetry points within the first Brillouin zone are represented in the inset of Figure 4(d). From the DOS, it is evident that there are significant contributions from Be $p$-orbitals, followed by Au $p$-orbital and so on. Because of the dominant role of Be $p$-orbital on the Fermi surface, the SOC, as well as the interaction effects are relatively weak, and the superconducting state is expected to be of phonon-mediated BCS type. The Fermi velocity plot in Figure 4(a) indicates that the large electron pocket at the $\Gamma$-point has significantly less Fermi velocity and large DOS. On the other hand, the DOS of the hole pocket at the $M$-point is significantly less, while the Fermi velocity is relatively larger. Therefore, it is rational to conclude that the pocket at the $\Gamma$-point leads to a larger superconducting gap ($\Delta_1$ $\simeq$ 320$\pm$10 $\mu$eV) while that at the $M$-point causes a smaller superconducting gap ($\Delta_2$ $\simeq$ 180$\pm$10 $\mu$eV). Hence, we have identified the bands that are responsible for multiband superconductivity in AuBe, and we have also found the distinct bands that give rise to the experimentally measured larger and smaller gaps, respectively.\\


In order to explain a detailed tunneling spectroscopic study on the vortex lattice of MgB$_2$ reported by Eskildsen et al.\cite{eskildsen2002vortex}, Koshelev and Golubov\cite{koshelev2003mixed} developed a general theory for a two-band superconductor under weak interband scattering. In such a superconductor, the two bands develop two different field scales, which can be revealed by the distributions of the order parameters and the local DOS. Consequently, two bands attain their normal state DOS at two different rates with respect to increasing magnetic field strength. Considering a negligible difference in the coupling constants, the ratio of the diffusion constants of the two bands is the only parameter to determine this relative rate. For MgB$_2$ as an example, the ratio 0.2 best explains the experimental results of Eskildsen \textit{et al.}\cite{eskildsen2002vortex}. To understand this effect in our present context for AuBe, we calculated the magnetic field dependence of the DOS for the respective bands using the formula for $N_{sj}(E)$ described earlier and the fitting parameters used in Figure 3(b). As it can be seen in Figure 4(b), the larger DOS $N_{s1}$ (corresponding to the electron pocket at the $\Gamma$-point and the larger gap $\Delta_1$) initially rises faster with increasing field to attain the normal state DOS (i.e. $N_{s}$=1) compared to the smaller DOS $N_{s2}$ (corresponding to the hole pocket at the M-point and the smaller gap $\Delta_2$). However, though $N_{s2}$ initially starts with a slower rate, latter it increases faster and attains the normal state DOS almost at the same field value $N_{s1}$ does. From visual interpretation, this result can be compared with the situation where the ratio of the diffusion constants is 1 in the proposed model. This suggests that, unlike in MgB$_2$, the two bands responsible for the superconductivity in AuBe have comparable transport characteristics. Possibly for the same reason, unlike in MgB$_2$, the two gaps are not distinctly resolved in the tunneling spectra of AuBe. To note, the model developed by Koshelev and Golubov\cite{koshelev2003mixed} had its own limitation as it assumed a large Ginzburg Landau parameter ($\kappa$). It serves the purpose for MgB$_2$ as $\kappa \gtrsim$ 10 for that material\cite{koshelev2003mixed}. However, in our present case for AuBe, reported $\kappa$ mostly varies between 0.4\cite{singh2019type} to 0.75\cite{rebar2019fermi} depending on the clean to dirty limit (except a widely different value 2.34 initially reported by \cite{amon2018noncentrosymmetric}). Some recent theoretical works\cite{silaev2011microscopic,vagov2016superconductivity} based on an intertype domain between type I and II can be interesting in this context and also relevant for multiband superconductors with low $\kappa$ like AuBe.\\


In summary, we have presented a detailed scanning tunneling spectroscopic investigation on non-centrosymmetric AuBe. From our analysis, we find that a single band model fails to explain the quasiparticle excitation data, but a simple two-band model provides a clean description of the same. From our two-gap analysis of tunneling spectra, we found that 2$\Delta_1/k_{B}T_{C}$ = 4.37 and 2$\Delta_2/k_{B}T_{C}$ = 2.46, respectively for the two bands. These values are consistent with the previous reports\cite{khasanov2020multiple,khasanov2020single}. Our magnetic field-dependent data support type-II behavior at sub-Kelvin temperature and show the survival of a finite local spectral gap far above the bulk critical field of AuBe. Our electronic band structure calculations, which are consistent with the previous reports\cite{amon2018noncentrosymmetric,rebar2019fermi}, support phonon-mediated BCS type superconductivity in this compound. Based on our Fermi surface analysis, we propose that the electron pocket at the $\Gamma$-point leads to the larger superconducting gap while the hole pocket at the $M$-point causes the smaller one.\\


R.P.S. would like to acknowledge financial support from the Science and Engineering Research Board (SERB)-Core Research Grant (grant No. CRG/2019/001028). G.S. acknowledges financial support from the Swarnajayanti fellowship awarded by the Department of Science and Technology (DST), Govt. of India (grant No. DST/SJF/PSA-01/2015-16).





\bibliography{Bibliography}	

\begin{thebibliography}{60}%
\makeatletter
\providecommand \@ifxundefined [1]{%
 \@ifx{#1\undefined}
}%
\providecommand \@ifnum [1]{%
 \ifnum #1\expandafter \@firstoftwo
 \else \expandafter \@secondoftwo
 \fi
}%
\providecommand \@ifx [1]{%
 \ifx #1\expandafter \@firstoftwo
 \else \expandafter \@secondoftwo
 \fi
}%
\providecommand \natexlab [1]{#1}%
\providecommand \enquote  [1]{``#1''}%
\providecommand \bibnamefont  [1]{#1}%
\providecommand \bibfnamefont [1]{#1}%
\providecommand \citenamefont [1]{#1}%
\providecommand \href@noop [0]{\@secondoftwo}%
\providecommand \href [0]{\begingroup \@sanitize@url \@href}%
\providecommand \@href[1]{\@@startlink{#1}\@@href}%
\providecommand \@@href[1]{\endgroup#1\@@endlink}%
\providecommand \@sanitize@url [0]{\catcode `\\12\catcode `\$12\catcode
  `\&12\catcode `\#12\catcode `\^12\catcode `\_12\catcode `\%12\relax}%
\providecommand \@@startlink[1]{}%
\providecommand \@@endlink[0]{}%
\providecommand \url  [0]{\begingroup\@sanitize@url \@url }%
\providecommand \@url [1]{\endgroup\@href {#1}{\urlprefix }}%
\providecommand \urlprefix  [0]{URL }%
\providecommand \Eprint [0]{\href }%
\providecommand \doibase [0]{https://doi.org/}%
\providecommand \selectlanguage [0]{\@gobble}%
\providecommand \bibinfo  [0]{\@secondoftwo}%
\providecommand \bibfield  [0]{\@secondoftwo}%
\providecommand \translation [1]{[#1]}%
\providecommand \BibitemOpen [0]{}%
\providecommand \bibitemStop [0]{}%
\providecommand \bibitemNoStop [0]{.\EOS\space}%
\providecommand \EOS [0]{\spacefactor3000\relax}%
\providecommand \BibitemShut  [1]{\csname bibitem#1\endcsname}%
\let\auto@bib@innerbib\@empty
\bibitem [{\citenamefont {Bauer}\ \emph {et~al.}(2010)\citenamefont {Bauer},
  \citenamefont {Rogl}, \citenamefont {Chen}, \citenamefont {Khan},
  \citenamefont {Michor}, \citenamefont {Hilscher}, \citenamefont {Royanian},
  \citenamefont {Kumagai}, \citenamefont {Li}, \citenamefont {Li} \emph
  {et~al.}}]{bauer2010unconventional}%
  \BibitemOpen
  \bibfield  {author} {\bibinfo {author} {\bibfnamefont {E.}~\bibnamefont
  {Bauer}}, \bibinfo {author} {\bibfnamefont {G.}~\bibnamefont {Rogl}},
  \bibinfo {author} {\bibfnamefont {X.-Q.}\ \bibnamefont {Chen}}, \bibinfo
  {author} {\bibfnamefont {R.}~\bibnamefont {Khan}}, \bibinfo {author}
  {\bibfnamefont {H.}~\bibnamefont {Michor}}, \bibinfo {author} {\bibfnamefont
  {G.}~\bibnamefont {Hilscher}}, \bibinfo {author} {\bibfnamefont
  {E.}~\bibnamefont {Royanian}}, \bibinfo {author} {\bibfnamefont
  {K.}~\bibnamefont {Kumagai}}, \bibinfo {author} {\bibfnamefont
  {D.}~\bibnamefont {Li}}, \bibinfo {author} {\bibfnamefont {Y.}~\bibnamefont
  {Li}}, \emph {et~al.},\ }\href@noop {} {\bibfield  {journal} {\bibinfo
  {journal} {Physical Review B}\ }\textbf {\bibinfo {volume} {82}},\ \bibinfo
  {pages} {064511} (\bibinfo {year} {2010})}\BibitemShut {NoStop}%
\bibitem [{\citenamefont {Mukuda}\ \emph {et~al.}(2008)\citenamefont {Mukuda},
  \citenamefont {Fujii}, \citenamefont {Ohara}, \citenamefont {Harada},
  \citenamefont {Yashima}, \citenamefont {Kitaoka}, \citenamefont {Okuda},
  \citenamefont {Settai},\ and\ \citenamefont {Onuki}}]{mukuda2008enhancement}%
  \BibitemOpen
  \bibfield  {author} {\bibinfo {author} {\bibfnamefont {H.}~\bibnamefont
  {Mukuda}}, \bibinfo {author} {\bibfnamefont {T.}~\bibnamefont {Fujii}},
  \bibinfo {author} {\bibfnamefont {T.}~\bibnamefont {Ohara}}, \bibinfo
  {author} {\bibfnamefont {A.}~\bibnamefont {Harada}}, \bibinfo {author}
  {\bibfnamefont {M.}~\bibnamefont {Yashima}}, \bibinfo {author} {\bibfnamefont
  {Y.}~\bibnamefont {Kitaoka}}, \bibinfo {author} {\bibfnamefont
  {Y.}~\bibnamefont {Okuda}}, \bibinfo {author} {\bibfnamefont
  {R.}~\bibnamefont {Settai}},\ and\ \bibinfo {author} {\bibfnamefont
  {Y.}~\bibnamefont {Onuki}},\ }\href@noop {} {\bibfield  {journal} {\bibinfo
  {journal} {Physical review letters}\ }\textbf {\bibinfo {volume} {100}},\
  \bibinfo {pages} {107003} (\bibinfo {year} {2008})}\BibitemShut {NoStop}%
\bibitem [{\citenamefont {Bonalde}\ \emph {et~al.}(2005)\citenamefont
  {Bonalde}, \citenamefont {Br{\"a}mer-Escamilla},\ and\ \citenamefont
  {Bauer}}]{bonalde2005evidence}%
  \BibitemOpen
  \bibfield  {author} {\bibinfo {author} {\bibfnamefont {I.}~\bibnamefont
  {Bonalde}}, \bibinfo {author} {\bibfnamefont {W.}~\bibnamefont
  {Br{\"a}mer-Escamilla}},\ and\ \bibinfo {author} {\bibfnamefont
  {E.}~\bibnamefont {Bauer}},\ }\href@noop {} {\bibfield  {journal} {\bibinfo
  {journal} {Physical review letters}\ }\textbf {\bibinfo {volume} {94}},\
  \bibinfo {pages} {207002} (\bibinfo {year} {2005})}\BibitemShut {NoStop}%
\bibitem [{\citenamefont {Frigeri}\ \emph {et~al.}(2004)\citenamefont
  {Frigeri}, \citenamefont {Agterberg}, \citenamefont {Koga},\ and\
  \citenamefont {Sigrist}}]{frigeri2004superconductivity}%
  \BibitemOpen
  \bibfield  {author} {\bibinfo {author} {\bibfnamefont {P.}~\bibnamefont
  {Frigeri}}, \bibinfo {author} {\bibfnamefont {D.}~\bibnamefont {Agterberg}},
  \bibinfo {author} {\bibfnamefont {A.}~\bibnamefont {Koga}},\ and\ \bibinfo
  {author} {\bibfnamefont {M.}~\bibnamefont {Sigrist}},\ }\href@noop {}
  {\bibfield  {journal} {\bibinfo  {journal} {Physical review letters}\
  }\textbf {\bibinfo {volume} {92}},\ \bibinfo {pages} {097001} (\bibinfo
  {year} {2004})}\BibitemShut {NoStop}%
\bibitem [{\citenamefont {Yuan}\ \emph {et~al.}(2006)\citenamefont {Yuan},
  \citenamefont {Agterberg}, \citenamefont {Hayashi}, \citenamefont {Badica},
  \citenamefont {Vandervelde}, \citenamefont {Togano}, \citenamefont
  {Sigrist},\ and\ \citenamefont {Salamon}}]{yuan2006s}%
  \BibitemOpen
  \bibfield  {author} {\bibinfo {author} {\bibfnamefont {H.}~\bibnamefont
  {Yuan}}, \bibinfo {author} {\bibfnamefont {D.}~\bibnamefont {Agterberg}},
  \bibinfo {author} {\bibfnamefont {N.}~\bibnamefont {Hayashi}}, \bibinfo
  {author} {\bibfnamefont {P.}~\bibnamefont {Badica}}, \bibinfo {author}
  {\bibfnamefont {D.}~\bibnamefont {Vandervelde}}, \bibinfo {author}
  {\bibfnamefont {K.}~\bibnamefont {Togano}}, \bibinfo {author} {\bibfnamefont
  {M.}~\bibnamefont {Sigrist}},\ and\ \bibinfo {author} {\bibfnamefont
  {M.}~\bibnamefont {Salamon}},\ }\href@noop {} {\bibfield  {journal} {\bibinfo
   {journal} {Physical review letters}\ }\textbf {\bibinfo {volume} {97}},\
  \bibinfo {pages} {017006} (\bibinfo {year} {2006})}\BibitemShut {NoStop}%
\bibitem [{\citenamefont {Nishiyama}\ \emph {et~al.}(2007)\citenamefont
  {Nishiyama}, \citenamefont {Inada},\ and\ \citenamefont
  {Zheng}}]{nishiyama2007spin}%
  \BibitemOpen
  \bibfield  {author} {\bibinfo {author} {\bibfnamefont {M.}~\bibnamefont
  {Nishiyama}}, \bibinfo {author} {\bibfnamefont {Y.}~\bibnamefont {Inada}},\
  and\ \bibinfo {author} {\bibfnamefont {G.-q.}\ \bibnamefont {Zheng}},\
  }\href@noop {} {\bibfield  {journal} {\bibinfo  {journal} {Physical review
  letters}\ }\textbf {\bibinfo {volume} {98}},\ \bibinfo {pages} {047002}
  (\bibinfo {year} {2007})}\BibitemShut {NoStop}%
\bibitem [{\citenamefont {Datta}\ \emph {et~al.}(2021)\citenamefont {Datta},
  \citenamefont {Vasdev}, \citenamefont {Ramachandran}, \citenamefont {Halder},
  \citenamefont {Motla}, \citenamefont {Kataria}, \citenamefont {Chowdhury},
  \citenamefont {Singh},\ and\ \citenamefont {Sheet}}]{datta2021spectroscopic}%
  \BibitemOpen
  \bibfield  {author} {\bibinfo {author} {\bibfnamefont {S.}~\bibnamefont
  {Datta}}, \bibinfo {author} {\bibfnamefont {A.}~\bibnamefont {Vasdev}},
  \bibinfo {author} {\bibfnamefont {R.}~\bibnamefont {Ramachandran}}, \bibinfo
  {author} {\bibfnamefont {S.}~\bibnamefont {Halder}}, \bibinfo {author}
  {\bibfnamefont {K.}~\bibnamefont {Motla}}, \bibinfo {author} {\bibfnamefont
  {A.}~\bibnamefont {Kataria}}, \bibinfo {author} {\bibfnamefont {R.~R.}\
  \bibnamefont {Chowdhury}}, \bibinfo {author} {\bibfnamefont {R.~P.}\
  \bibnamefont {Singh}},\ and\ \bibinfo {author} {\bibfnamefont
  {G.}~\bibnamefont {Sheet}},\ }\href@noop {} {\bibfield  {journal} {\bibinfo
  {journal} {Scientific Reports}\ }\textbf {\bibinfo {volume} {11}},\ \bibinfo
  {pages} {21030} (\bibinfo {year} {2021})}\BibitemShut {NoStop}%
\bibitem [{\citenamefont {Karki}\ \emph {et~al.}(2011)\citenamefont {Karki},
  \citenamefont {Xiong}, \citenamefont {Haldolaarachchige}, \citenamefont
  {Stadler}, \citenamefont {Vekhter}, \citenamefont {Adams}, \citenamefont
  {Young}, \citenamefont {Phelan},\ and\ \citenamefont
  {Chan}}]{karki2011physical}%
  \BibitemOpen
  \bibfield  {author} {\bibinfo {author} {\bibfnamefont {A.~B.}\ \bibnamefont
  {Karki}}, \bibinfo {author} {\bibfnamefont {Y.~M.}\ \bibnamefont {Xiong}},
  \bibinfo {author} {\bibfnamefont {N.}~\bibnamefont {Haldolaarachchige}},
  \bibinfo {author} {\bibfnamefont {S.}~\bibnamefont {Stadler}}, \bibinfo
  {author} {\bibfnamefont {I.}~\bibnamefont {Vekhter}}, \bibinfo {author}
  {\bibfnamefont {P.~W.}\ \bibnamefont {Adams}}, \bibinfo {author}
  {\bibfnamefont {D.}~\bibnamefont {Young}}, \bibinfo {author} {\bibfnamefont
  {W.}~\bibnamefont {Phelan}},\ and\ \bibinfo {author} {\bibfnamefont {J.~Y.}\
  \bibnamefont {Chan}},\ }\href@noop {} {\bibfield  {journal} {\bibinfo
  {journal} {Physical Review B}\ }\textbf {\bibinfo {volume} {83}},\ \bibinfo
  {pages} {144525} (\bibinfo {year} {2011})}\BibitemShut {NoStop}%
\bibitem [{\citenamefont {Sun}\ \emph {et~al.}(2015)\citenamefont {Sun},
  \citenamefont {Enayat}, \citenamefont {Maldonado}, \citenamefont {Lithgow},
  \citenamefont {Yelland}, \citenamefont {Peets}, \citenamefont {Yaresko},
  \citenamefont {Schnyder},\ and\ \citenamefont {Wahl}}]{sun2015dirac}%
  \BibitemOpen
  \bibfield  {author} {\bibinfo {author} {\bibfnamefont {Z.}~\bibnamefont
  {Sun}}, \bibinfo {author} {\bibfnamefont {M.}~\bibnamefont {Enayat}},
  \bibinfo {author} {\bibfnamefont {A.}~\bibnamefont {Maldonado}}, \bibinfo
  {author} {\bibfnamefont {C.}~\bibnamefont {Lithgow}}, \bibinfo {author}
  {\bibfnamefont {E.}~\bibnamefont {Yelland}}, \bibinfo {author} {\bibfnamefont
  {D.~C.}\ \bibnamefont {Peets}}, \bibinfo {author} {\bibfnamefont
  {A.}~\bibnamefont {Yaresko}}, \bibinfo {author} {\bibfnamefont {A.~P.}\
  \bibnamefont {Schnyder}},\ and\ \bibinfo {author} {\bibfnamefont
  {P.}~\bibnamefont {Wahl}},\ }\href@noop {} {\bibfield  {journal} {\bibinfo
  {journal} {Nature communications}\ }\textbf {\bibinfo {volume} {6}},\
  \bibinfo {pages} {1} (\bibinfo {year} {2015})}\BibitemShut {NoStop}%
\bibitem [{\citenamefont {Yan}\ \emph {et~al.}(2016)\citenamefont {Yan},
  \citenamefont {Xu}, \citenamefont {He}, \citenamefont {Dong}, \citenamefont
  {Cho}, \citenamefont {Peets}, \citenamefont {Park},\ and\ \citenamefont
  {Li}}]{yan2016nodeless}%
  \BibitemOpen
  \bibfield  {author} {\bibinfo {author} {\bibfnamefont {X.}~\bibnamefont
  {Yan}}, \bibinfo {author} {\bibfnamefont {Y.}~\bibnamefont {Xu}}, \bibinfo
  {author} {\bibfnamefont {L.}~\bibnamefont {He}}, \bibinfo {author}
  {\bibfnamefont {J.}~\bibnamefont {Dong}}, \bibinfo {author} {\bibfnamefont
  {H.}~\bibnamefont {Cho}}, \bibinfo {author} {\bibfnamefont {D.~C.}\
  \bibnamefont {Peets}}, \bibinfo {author} {\bibfnamefont {J.-G.}\ \bibnamefont
  {Park}},\ and\ \bibinfo {author} {\bibfnamefont {S.}~\bibnamefont {Li}},\
  }\href@noop {} {\bibfield  {journal} {\bibinfo  {journal} {Superconductor
  Science and Technology}\ }\textbf {\bibinfo {volume} {29}},\ \bibinfo {pages}
  {065001} (\bibinfo {year} {2016})}\BibitemShut {NoStop}%
\bibitem [{\citenamefont {Shibayama}\ \emph {et~al.}(2007)\citenamefont
  {Shibayama}, \citenamefont {Nohara}, \citenamefont {Katori}, \citenamefont
  {Okamoto}, \citenamefont {Hiroi},\ and\ \citenamefont
  {Takagi}}]{shibayama2007superconductivity}%
  \BibitemOpen
  \bibfield  {author} {\bibinfo {author} {\bibfnamefont {T.}~\bibnamefont
  {Shibayama}}, \bibinfo {author} {\bibfnamefont {M.}~\bibnamefont {Nohara}},
  \bibinfo {author} {\bibfnamefont {H.~A.}\ \bibnamefont {Katori}}, \bibinfo
  {author} {\bibfnamefont {Y.}~\bibnamefont {Okamoto}}, \bibinfo {author}
  {\bibfnamefont {Z.}~\bibnamefont {Hiroi}},\ and\ \bibinfo {author}
  {\bibfnamefont {H.}~\bibnamefont {Takagi}},\ }\href@noop {} {\bibfield
  {journal} {\bibinfo  {journal} {Journal of the Physical Society of Japan}\
  }\textbf {\bibinfo {volume} {76}},\ \bibinfo {pages} {073708} (\bibinfo
  {year} {2007})}\BibitemShut {NoStop}%
\bibitem [{\citenamefont {Wakui}\ \emph {et~al.}(2009)\citenamefont {Wakui},
  \citenamefont {Akutagawa}, \citenamefont {Kase}, \citenamefont {Kawashima},
  \citenamefont {Muranaka}, \citenamefont {Iwahori}, \citenamefont {Abe},\ and\
  \citenamefont {Akimitsu}}]{wakui2009thermodynamic}%
  \BibitemOpen
  \bibfield  {author} {\bibinfo {author} {\bibfnamefont {K.}~\bibnamefont
  {Wakui}}, \bibinfo {author} {\bibfnamefont {S.}~\bibnamefont {Akutagawa}},
  \bibinfo {author} {\bibfnamefont {N.}~\bibnamefont {Kase}}, \bibinfo {author}
  {\bibfnamefont {K.}~\bibnamefont {Kawashima}}, \bibinfo {author}
  {\bibfnamefont {T.}~\bibnamefont {Muranaka}}, \bibinfo {author}
  {\bibfnamefont {Y.}~\bibnamefont {Iwahori}}, \bibinfo {author} {\bibfnamefont
  {J.}~\bibnamefont {Abe}},\ and\ \bibinfo {author} {\bibfnamefont
  {J.}~\bibnamefont {Akimitsu}},\ }\href@noop {} {\bibfield  {journal}
  {\bibinfo  {journal} {journal of the physical society of japan}\ }\textbf
  {\bibinfo {volume} {78}},\ \bibinfo {pages} {034710} (\bibinfo {year}
  {2009})}\BibitemShut {NoStop}%
\bibitem [{\citenamefont {Anand}\ \emph {et~al.}(2011)\citenamefont {Anand},
  \citenamefont {Hillier}, \citenamefont {Adroja}, \citenamefont {Strydom},
  \citenamefont {Michor}, \citenamefont {McEwen},\ and\ \citenamefont
  {Rainford}}]{anand2011specific}%
  \BibitemOpen
  \bibfield  {author} {\bibinfo {author} {\bibfnamefont {V.}~\bibnamefont
  {Anand}}, \bibinfo {author} {\bibfnamefont {A.}~\bibnamefont {Hillier}},
  \bibinfo {author} {\bibfnamefont {D.}~\bibnamefont {Adroja}}, \bibinfo
  {author} {\bibfnamefont {A.}~\bibnamefont {Strydom}}, \bibinfo {author}
  {\bibfnamefont {H.}~\bibnamefont {Michor}}, \bibinfo {author} {\bibfnamefont
  {K.}~\bibnamefont {McEwen}},\ and\ \bibinfo {author} {\bibfnamefont
  {B.}~\bibnamefont {Rainford}},\ }\href@noop {} {\bibfield  {journal}
  {\bibinfo  {journal} {Physical Review B}\ }\textbf {\bibinfo {volume} {83}},\
  \bibinfo {pages} {064522} (\bibinfo {year} {2011})}\BibitemShut {NoStop}%
\bibitem [{\citenamefont {Anand}\ \emph {et~al.}(2014)\citenamefont {Anand},
  \citenamefont {Britz}, \citenamefont {Bhattacharyya}, \citenamefont {Adroja},
  \citenamefont {Hillier}, \citenamefont {Strydom}, \citenamefont {Kockelmann},
  \citenamefont {Rainford},\ and\ \citenamefont {McEwen}}]{anand2014physical}%
  \BibitemOpen
  \bibfield  {author} {\bibinfo {author} {\bibfnamefont {V.~K.}\ \bibnamefont
  {Anand}}, \bibinfo {author} {\bibfnamefont {D.}~\bibnamefont {Britz}},
  \bibinfo {author} {\bibfnamefont {A.}~\bibnamefont {Bhattacharyya}}, \bibinfo
  {author} {\bibfnamefont {D.}~\bibnamefont {Adroja}}, \bibinfo {author}
  {\bibfnamefont {A.}~\bibnamefont {Hillier}}, \bibinfo {author} {\bibfnamefont
  {A.}~\bibnamefont {Strydom}}, \bibinfo {author} {\bibfnamefont
  {W.}~\bibnamefont {Kockelmann}}, \bibinfo {author} {\bibfnamefont
  {B.}~\bibnamefont {Rainford}},\ and\ \bibinfo {author} {\bibfnamefont
  {K.~A.}\ \bibnamefont {McEwen}},\ }\href@noop {} {\bibfield  {journal}
  {\bibinfo  {journal} {Physical Review B}\ }\textbf {\bibinfo {volume} {90}},\
  \bibinfo {pages} {014513} (\bibinfo {year} {2014})}\BibitemShut {NoStop}%
\bibitem [{\citenamefont {Smidman}\ \emph {et~al.}(2014)\citenamefont
  {Smidman}, \citenamefont {Hillier}, \citenamefont {Adroja}, \citenamefont
  {Lees}, \citenamefont {Anand}, \citenamefont {Singh}, \citenamefont {Smith},
  \citenamefont {Paul},\ and\ \citenamefont
  {Balakrishnan}}]{smidman2014investigations}%
  \BibitemOpen
  \bibfield  {author} {\bibinfo {author} {\bibfnamefont {M.}~\bibnamefont
  {Smidman}}, \bibinfo {author} {\bibfnamefont {A.}~\bibnamefont {Hillier}},
  \bibinfo {author} {\bibfnamefont {D.}~\bibnamefont {Adroja}}, \bibinfo
  {author} {\bibfnamefont {M.}~\bibnamefont {Lees}}, \bibinfo {author}
  {\bibfnamefont {V.~K.}\ \bibnamefont {Anand}}, \bibinfo {author}
  {\bibfnamefont {R.~P.}\ \bibnamefont {Singh}}, \bibinfo {author}
  {\bibfnamefont {R.}~\bibnamefont {Smith}}, \bibinfo {author} {\bibfnamefont
  {D.}~\bibnamefont {Paul}},\ and\ \bibinfo {author} {\bibfnamefont
  {G.}~\bibnamefont {Balakrishnan}},\ }\href@noop {} {\bibfield  {journal}
  {\bibinfo  {journal} {Physical Review B}\ }\textbf {\bibinfo {volume} {89}},\
  \bibinfo {pages} {094509} (\bibinfo {year} {2014})}\BibitemShut {NoStop}%
\bibitem [{\citenamefont {Chen}\ \emph {et~al.}(2013)\citenamefont {Chen},
  \citenamefont {Jiao}, \citenamefont {Zhang}, \citenamefont {Chen},
  \citenamefont {Yang}, \citenamefont {Nicklas}, \citenamefont {Steglich},\
  and\ \citenamefont {Yuan}}]{chen2013evidence}%
  \BibitemOpen
  \bibfield  {author} {\bibinfo {author} {\bibfnamefont {J.}~\bibnamefont
  {Chen}}, \bibinfo {author} {\bibfnamefont {L.}~\bibnamefont {Jiao}}, \bibinfo
  {author} {\bibfnamefont {J.}~\bibnamefont {Zhang}}, \bibinfo {author}
  {\bibfnamefont {Y.}~\bibnamefont {Chen}}, \bibinfo {author} {\bibfnamefont
  {L.}~\bibnamefont {Yang}}, \bibinfo {author} {\bibfnamefont {M.}~\bibnamefont
  {Nicklas}}, \bibinfo {author} {\bibfnamefont {F.}~\bibnamefont {Steglich}},\
  and\ \bibinfo {author} {\bibfnamefont {H.}~\bibnamefont {Yuan}},\ }\href@noop
  {} {\bibfield  {journal} {\bibinfo  {journal} {New Journal of Physics}\
  }\textbf {\bibinfo {volume} {15}},\ \bibinfo {pages} {053005} (\bibinfo
  {year} {2013})}\BibitemShut {NoStop}%
\bibitem [{\citenamefont {Mayoh}\ \emph {et~al.}(2018)\citenamefont {Mayoh},
  \citenamefont {Hillier}, \citenamefont {G{\"o}tze}, \citenamefont {Paul},
  \citenamefont {Balakrishnan},\ and\ \citenamefont
  {Lees}}]{mayoh2018multigap}%
  \BibitemOpen
  \bibfield  {author} {\bibinfo {author} {\bibfnamefont {D.}~\bibnamefont
  {Mayoh}}, \bibinfo {author} {\bibfnamefont {A.}~\bibnamefont {Hillier}},
  \bibinfo {author} {\bibfnamefont {K.}~\bibnamefont {G{\"o}tze}}, \bibinfo
  {author} {\bibfnamefont {D.~M.}\ \bibnamefont {Paul}}, \bibinfo {author}
  {\bibfnamefont {G.}~\bibnamefont {Balakrishnan}},\ and\ \bibinfo {author}
  {\bibfnamefont {M.}~\bibnamefont {Lees}},\ }\href@noop {} {\bibfield
  {journal} {\bibinfo  {journal} {Physical Review B}\ }\textbf {\bibinfo
  {volume} {98}},\ \bibinfo {pages} {014502} (\bibinfo {year}
  {2018})}\BibitemShut {NoStop}%
\bibitem [{\citenamefont {Sugawara}\ \emph {et~al.}(2007)\citenamefont
  {Sugawara}, \citenamefont {Sato}, \citenamefont {Souma}, \citenamefont
  {Takahashi},\ and\ \citenamefont {Ochiai}}]{sugawara2007anomalous}%
  \BibitemOpen
  \bibfield  {author} {\bibinfo {author} {\bibfnamefont {K.}~\bibnamefont
  {Sugawara}}, \bibinfo {author} {\bibfnamefont {T.}~\bibnamefont {Sato}},
  \bibinfo {author} {\bibfnamefont {S.}~\bibnamefont {Souma}}, \bibinfo
  {author} {\bibfnamefont {T.}~\bibnamefont {Takahashi}},\ and\ \bibinfo
  {author} {\bibfnamefont {A.}~\bibnamefont {Ochiai}},\ }\href@noop {}
  {\bibfield  {journal} {\bibinfo  {journal} {Physical Review B}\ }\textbf
  {\bibinfo {volume} {76}},\ \bibinfo {pages} {132512} (\bibinfo {year}
  {2007})}\BibitemShut {NoStop}%
\bibitem [{\citenamefont {Kuroiwa}\ \emph {et~al.}(2008)\citenamefont
  {Kuroiwa}, \citenamefont {Saura}, \citenamefont {Akimitsu}, \citenamefont
  {Hiraishi}, \citenamefont {Miyazaki}, \citenamefont {Satoh}, \citenamefont
  {Takeshita},\ and\ \citenamefont {Kadono}}]{kuroiwa2008multigap}%
  \BibitemOpen
  \bibfield  {author} {\bibinfo {author} {\bibfnamefont {S.}~\bibnamefont
  {Kuroiwa}}, \bibinfo {author} {\bibfnamefont {Y.}~\bibnamefont {Saura}},
  \bibinfo {author} {\bibfnamefont {J.}~\bibnamefont {Akimitsu}}, \bibinfo
  {author} {\bibfnamefont {M.}~\bibnamefont {Hiraishi}}, \bibinfo {author}
  {\bibfnamefont {M.}~\bibnamefont {Miyazaki}}, \bibinfo {author}
  {\bibfnamefont {K.}~\bibnamefont {Satoh}}, \bibinfo {author} {\bibfnamefont
  {S.}~\bibnamefont {Takeshita}},\ and\ \bibinfo {author} {\bibfnamefont
  {R.}~\bibnamefont {Kadono}},\ }\href@noop {} {\bibfield  {journal} {\bibinfo
  {journal} {Physical review letters}\ }\textbf {\bibinfo {volume} {100}},\
  \bibinfo {pages} {097002} (\bibinfo {year} {2008})}\BibitemShut {NoStop}%
\bibitem [{\citenamefont {Matthias}(1959)}]{matthias1959superconductivity}%
  \BibitemOpen
  \bibfield  {author} {\bibinfo {author} {\bibfnamefont {B.}~\bibnamefont
  {Matthias}},\ }\href@noop {} {\bibfield  {journal} {\bibinfo  {journal}
  {Journal of Physics and Chemistry of Solids}\ }\textbf {\bibinfo {volume}
  {10}},\ \bibinfo {pages} {342} (\bibinfo {year} {1959})}\BibitemShut
  {NoStop}%
\bibitem [{\citenamefont {Rebar}(2015)}]{rebar2015exploring}%
  \BibitemOpen
  \bibfield  {author} {\bibinfo {author} {\bibfnamefont {D.~J.}\ \bibnamefont
  {Rebar}},\ }\href@noop {} {\  (\bibinfo {year} {2015})}\BibitemShut {NoStop}%
\bibitem [{\citenamefont {Amon}\ \emph {et~al.}(2018)\citenamefont {Amon},
  \citenamefont {Svanidze}, \citenamefont {Cardoso-Gil}, \citenamefont
  {Wilson}, \citenamefont {Rosner}, \citenamefont {Bobnar}, \citenamefont
  {Schnelle}, \citenamefont {Lynn}, \citenamefont {Gumeniuk}, \citenamefont
  {Hennig} \emph {et~al.}}]{amon2018noncentrosymmetric}%
  \BibitemOpen
  \bibfield  {author} {\bibinfo {author} {\bibfnamefont {A.}~\bibnamefont
  {Amon}}, \bibinfo {author} {\bibfnamefont {E.}~\bibnamefont {Svanidze}},
  \bibinfo {author} {\bibfnamefont {R.}~\bibnamefont {Cardoso-Gil}}, \bibinfo
  {author} {\bibfnamefont {M.}~\bibnamefont {Wilson}}, \bibinfo {author}
  {\bibfnamefont {H.}~\bibnamefont {Rosner}}, \bibinfo {author} {\bibfnamefont
  {M.}~\bibnamefont {Bobnar}}, \bibinfo {author} {\bibfnamefont
  {W.}~\bibnamefont {Schnelle}}, \bibinfo {author} {\bibfnamefont {J.~W.}\
  \bibnamefont {Lynn}}, \bibinfo {author} {\bibfnamefont {R.}~\bibnamefont
  {Gumeniuk}}, \bibinfo {author} {\bibfnamefont {C.}~\bibnamefont {Hennig}},
  \emph {et~al.},\ }\href@noop {} {\bibfield  {journal} {\bibinfo  {journal}
  {Physical Review B}\ }\textbf {\bibinfo {volume} {97}},\ \bibinfo {pages}
  {014501} (\bibinfo {year} {2018})}\BibitemShut {NoStop}%
\bibitem [{\citenamefont {Singh}\ \emph {et~al.}(2019)\citenamefont {Singh},
  \citenamefont {Hillier},\ and\ \citenamefont {Singh}}]{singh2019type}%
  \BibitemOpen
  \bibfield  {author} {\bibinfo {author} {\bibfnamefont {D.}~\bibnamefont
  {Singh}}, \bibinfo {author} {\bibfnamefont {A.}~\bibnamefont {Hillier}},\
  and\ \bibinfo {author} {\bibfnamefont {R.}~\bibnamefont {Singh}},\
  }\href@noop {} {\bibfield  {journal} {\bibinfo  {journal} {Physical Review
  B}\ }\textbf {\bibinfo {volume} {99}},\ \bibinfo {pages} {134509} (\bibinfo
  {year} {2019})}\BibitemShut {NoStop}%
\bibitem [{\citenamefont {Beare}\ \emph {et~al.}(2019)\citenamefont {Beare},
  \citenamefont {Nugent}, \citenamefont {Wilson}, \citenamefont {Cai},
  \citenamefont {Munsie}, \citenamefont {Amon}, \citenamefont {Leithe-Jasper},
  \citenamefont {Gong}, \citenamefont {Guo}, \citenamefont {Guguchia} \emph
  {et~al.}}]{beare2019mu}%
  \BibitemOpen
  \bibfield  {author} {\bibinfo {author} {\bibfnamefont {J.}~\bibnamefont
  {Beare}}, \bibinfo {author} {\bibfnamefont {M.}~\bibnamefont {Nugent}},
  \bibinfo {author} {\bibfnamefont {M.}~\bibnamefont {Wilson}}, \bibinfo
  {author} {\bibfnamefont {Y.}~\bibnamefont {Cai}}, \bibinfo {author}
  {\bibfnamefont {T.}~\bibnamefont {Munsie}}, \bibinfo {author} {\bibfnamefont
  {A.}~\bibnamefont {Amon}}, \bibinfo {author} {\bibfnamefont {A.}~\bibnamefont
  {Leithe-Jasper}}, \bibinfo {author} {\bibfnamefont {Z.}~\bibnamefont {Gong}},
  \bibinfo {author} {\bibfnamefont {S.}~\bibnamefont {Guo}}, \bibinfo {author}
  {\bibfnamefont {Z.}~\bibnamefont {Guguchia}}, \emph {et~al.},\ }\href@noop {}
  {\bibfield  {journal} {\bibinfo  {journal} {Physical Review B}\ }\textbf
  {\bibinfo {volume} {99}},\ \bibinfo {pages} {134510} (\bibinfo {year}
  {2019})}\BibitemShut {NoStop}%
\bibitem [{\citenamefont {Rebar}\ \emph {et~al.}(2019)\citenamefont {Rebar},
  \citenamefont {Birnbaum}, \citenamefont {Singleton}, \citenamefont {Khan},
  \citenamefont {Ball}, \citenamefont {Adams}, \citenamefont {Chan},
  \citenamefont {Young}, \citenamefont {Browne},\ and\ \citenamefont
  {DiTusa}}]{rebar2019fermi}%
  \BibitemOpen
  \bibfield  {author} {\bibinfo {author} {\bibfnamefont {D.~J.}\ \bibnamefont
  {Rebar}}, \bibinfo {author} {\bibfnamefont {S.~M.}\ \bibnamefont {Birnbaum}},
  \bibinfo {author} {\bibfnamefont {J.}~\bibnamefont {Singleton}}, \bibinfo
  {author} {\bibfnamefont {M.}~\bibnamefont {Khan}}, \bibinfo {author}
  {\bibfnamefont {J.}~\bibnamefont {Ball}}, \bibinfo {author} {\bibfnamefont
  {P.}~\bibnamefont {Adams}}, \bibinfo {author} {\bibfnamefont {J.~Y.}\
  \bibnamefont {Chan}}, \bibinfo {author} {\bibfnamefont {D.}~\bibnamefont
  {Young}}, \bibinfo {author} {\bibfnamefont {D.~A.}\ \bibnamefont {Browne}},\
  and\ \bibinfo {author} {\bibfnamefont {J.~F.}\ \bibnamefont {DiTusa}},\
  }\href@noop {} {\bibfield  {journal} {\bibinfo  {journal} {Physical Review
  B}\ }\textbf {\bibinfo {volume} {99}},\ \bibinfo {pages} {094517} (\bibinfo
  {year} {2019})}\BibitemShut {NoStop}%
\bibitem [{\citenamefont {Bradlyn}\ \emph {et~al.}(2016)\citenamefont
  {Bradlyn}, \citenamefont {Cano}, \citenamefont {Wang}, \citenamefont
  {Vergniory}, \citenamefont {Felser}, \citenamefont {Cava},\ and\
  \citenamefont {Bernevig}}]{bradlyn2016beyond}%
  \BibitemOpen
  \bibfield  {author} {\bibinfo {author} {\bibfnamefont {B.}~\bibnamefont
  {Bradlyn}}, \bibinfo {author} {\bibfnamefont {J.}~\bibnamefont {Cano}},
  \bibinfo {author} {\bibfnamefont {Z.}~\bibnamefont {Wang}}, \bibinfo {author}
  {\bibfnamefont {M.}~\bibnamefont {Vergniory}}, \bibinfo {author}
  {\bibfnamefont {C.}~\bibnamefont {Felser}}, \bibinfo {author} {\bibfnamefont
  {R.~J.}\ \bibnamefont {Cava}},\ and\ \bibinfo {author} {\bibfnamefont
  {B.~A.}\ \bibnamefont {Bernevig}},\ }\href@noop {} {\bibfield  {journal}
  {\bibinfo  {journal} {Science}\ }\textbf {\bibinfo {volume} {353}} (\bibinfo
  {year} {2016})}\BibitemShut {NoStop}%
\bibitem [{\citenamefont {Tang}\ \emph {et~al.}(2017)\citenamefont {Tang},
  \citenamefont {Zhou},\ and\ \citenamefont {Zhang}}]{tang2017multiple}%
  \BibitemOpen
  \bibfield  {author} {\bibinfo {author} {\bibfnamefont {P.}~\bibnamefont
  {Tang}}, \bibinfo {author} {\bibfnamefont {Q.}~\bibnamefont {Zhou}},\ and\
  \bibinfo {author} {\bibfnamefont {S.-C.}\ \bibnamefont {Zhang}},\ }\href@noop
  {} {\bibfield  {journal} {\bibinfo  {journal} {Physical review letters}\
  }\textbf {\bibinfo {volume} {119}},\ \bibinfo {pages} {206402} (\bibinfo
  {year} {2017})}\BibitemShut {NoStop}%
\bibitem [{\citenamefont {Chang}\ \emph {et~al.}(2017)\citenamefont {Chang},
  \citenamefont {Xu}, \citenamefont {Wieder}, \citenamefont {Sanchez},
  \citenamefont {Huang}, \citenamefont {Belopolski}, \citenamefont {Chang},
  \citenamefont {Zhang}, \citenamefont {Bansil}, \citenamefont {Lin} \emph
  {et~al.}}]{chang2017unconventional}%
  \BibitemOpen
  \bibfield  {author} {\bibinfo {author} {\bibfnamefont {G.}~\bibnamefont
  {Chang}}, \bibinfo {author} {\bibfnamefont {S.-Y.}\ \bibnamefont {Xu}},
  \bibinfo {author} {\bibfnamefont {B.~J.}\ \bibnamefont {Wieder}}, \bibinfo
  {author} {\bibfnamefont {D.~S.}\ \bibnamefont {Sanchez}}, \bibinfo {author}
  {\bibfnamefont {S.-M.}\ \bibnamefont {Huang}}, \bibinfo {author}
  {\bibfnamefont {I.}~\bibnamefont {Belopolski}}, \bibinfo {author}
  {\bibfnamefont {T.-R.}\ \bibnamefont {Chang}}, \bibinfo {author}
  {\bibfnamefont {S.}~\bibnamefont {Zhang}}, \bibinfo {author} {\bibfnamefont
  {A.}~\bibnamefont {Bansil}}, \bibinfo {author} {\bibfnamefont
  {H.}~\bibnamefont {Lin}}, \emph {et~al.},\ }\href@noop {} {\bibfield
  {journal} {\bibinfo  {journal} {Physical review letters}\ }\textbf {\bibinfo
  {volume} {119}},\ \bibinfo {pages} {206401} (\bibinfo {year}
  {2017})}\BibitemShut {NoStop}%
\bibitem [{\citenamefont {Yu}\ \emph {et~al.}(2011)\citenamefont {Yu},
  \citenamefont {Kanazawa}, \citenamefont {Onose}, \citenamefont {Kimoto},
  \citenamefont {Zhang}, \citenamefont {Ishiwata}, \citenamefont {Matsui},\
  and\ \citenamefont {Tokura}}]{yu2011near}%
  \BibitemOpen
  \bibfield  {author} {\bibinfo {author} {\bibfnamefont {X.}~\bibnamefont
  {Yu}}, \bibinfo {author} {\bibfnamefont {N.}~\bibnamefont {Kanazawa}},
  \bibinfo {author} {\bibfnamefont {Y.}~\bibnamefont {Onose}}, \bibinfo
  {author} {\bibfnamefont {K.}~\bibnamefont {Kimoto}}, \bibinfo {author}
  {\bibfnamefont {W.}~\bibnamefont {Zhang}}, \bibinfo {author} {\bibfnamefont
  {S.}~\bibnamefont {Ishiwata}}, \bibinfo {author} {\bibfnamefont
  {Y.}~\bibnamefont {Matsui}},\ and\ \bibinfo {author} {\bibfnamefont
  {Y.}~\bibnamefont {Tokura}},\ }\href@noop {} {\bibfield  {journal} {\bibinfo
  {journal} {Nature materials}\ }\textbf {\bibinfo {volume} {10}},\ \bibinfo
  {pages} {106} (\bibinfo {year} {2011})}\BibitemShut {NoStop}%
\bibitem [{\citenamefont {Tonomura}\ \emph {et~al.}(2012)\citenamefont
  {Tonomura}, \citenamefont {Yu}, \citenamefont {Yanagisawa}, \citenamefont
  {Matsuda}, \citenamefont {Onose}, \citenamefont {Kanazawa}, \citenamefont
  {Park},\ and\ \citenamefont {Tokura}}]{tonomura2012real}%
  \BibitemOpen
  \bibfield  {author} {\bibinfo {author} {\bibfnamefont {A.}~\bibnamefont
  {Tonomura}}, \bibinfo {author} {\bibfnamefont {X.}~\bibnamefont {Yu}},
  \bibinfo {author} {\bibfnamefont {K.}~\bibnamefont {Yanagisawa}}, \bibinfo
  {author} {\bibfnamefont {T.}~\bibnamefont {Matsuda}}, \bibinfo {author}
  {\bibfnamefont {Y.}~\bibnamefont {Onose}}, \bibinfo {author} {\bibfnamefont
  {N.}~\bibnamefont {Kanazawa}}, \bibinfo {author} {\bibfnamefont {H.~S.}\
  \bibnamefont {Park}},\ and\ \bibinfo {author} {\bibfnamefont
  {Y.}~\bibnamefont {Tokura}},\ }\href@noop {} {\bibfield  {journal} {\bibinfo
  {journal} {Nano letters}\ }\textbf {\bibinfo {volume} {12}},\ \bibinfo
  {pages} {1673} (\bibinfo {year} {2012})}\BibitemShut {NoStop}%
\bibitem [{\citenamefont {Kanazawa}\ \emph {et~al.}(2012)\citenamefont
  {Kanazawa}, \citenamefont {Kim}, \citenamefont {Inosov}, \citenamefont
  {White}, \citenamefont {Egetenmeyer}, \citenamefont {Gavilano}, \citenamefont
  {Ishiwata}, \citenamefont {Onose}, \citenamefont {Arima}, \citenamefont
  {Keimer} \emph {et~al.}}]{kanazawa2012possible}%
  \BibitemOpen
  \bibfield  {author} {\bibinfo {author} {\bibfnamefont {N.}~\bibnamefont
  {Kanazawa}}, \bibinfo {author} {\bibfnamefont {J.-H.}\ \bibnamefont {Kim}},
  \bibinfo {author} {\bibfnamefont {D.}~\bibnamefont {Inosov}}, \bibinfo
  {author} {\bibfnamefont {J.}~\bibnamefont {White}}, \bibinfo {author}
  {\bibfnamefont {N.}~\bibnamefont {Egetenmeyer}}, \bibinfo {author}
  {\bibfnamefont {J.}~\bibnamefont {Gavilano}}, \bibinfo {author}
  {\bibfnamefont {S.}~\bibnamefont {Ishiwata}}, \bibinfo {author}
  {\bibfnamefont {Y.}~\bibnamefont {Onose}}, \bibinfo {author} {\bibfnamefont
  {T.}~\bibnamefont {Arima}}, \bibinfo {author} {\bibfnamefont
  {B.}~\bibnamefont {Keimer}}, \emph {et~al.},\ }\href@noop {} {\bibfield
  {journal} {\bibinfo  {journal} {Physical Review B}\ }\textbf {\bibinfo
  {volume} {86}},\ \bibinfo {pages} {134425} (\bibinfo {year}
  {2012})}\BibitemShut {NoStop}%
\bibitem [{\citenamefont {Khasanov}\ \emph
  {et~al.}(2020{\natexlab{a}})\citenamefont {Khasanov}, \citenamefont {Gupta},
  \citenamefont {Das}, \citenamefont {Amon}, \citenamefont {Leithe-Jasper},\
  and\ \citenamefont {Svanidze}}]{khasanov2020multiple}%
  \BibitemOpen
  \bibfield  {author} {\bibinfo {author} {\bibfnamefont {R.}~\bibnamefont
  {Khasanov}}, \bibinfo {author} {\bibfnamefont {R.}~\bibnamefont {Gupta}},
  \bibinfo {author} {\bibfnamefont {D.}~\bibnamefont {Das}}, \bibinfo {author}
  {\bibfnamefont {A.}~\bibnamefont {Amon}}, \bibinfo {author} {\bibfnamefont
  {A.}~\bibnamefont {Leithe-Jasper}},\ and\ \bibinfo {author} {\bibfnamefont
  {E.}~\bibnamefont {Svanidze}},\ }\href@noop {} {\bibfield  {journal}
  {\bibinfo  {journal} {Physical Review Research}\ }\textbf {\bibinfo {volume}
  {2}},\ \bibinfo {pages} {023142} (\bibinfo {year}
  {2020}{\natexlab{a}})}\BibitemShut {NoStop}%
\bibitem [{\citenamefont {Khasanov}\ \emph
  {et~al.}(2020{\natexlab{b}})\citenamefont {Khasanov}, \citenamefont {Gupta},
  \citenamefont {Das}, \citenamefont {Leithe-Jasper},\ and\ \citenamefont
  {Svanidze}}]{khasanov2020single}%
  \BibitemOpen
  \bibfield  {author} {\bibinfo {author} {\bibfnamefont {R.}~\bibnamefont
  {Khasanov}}, \bibinfo {author} {\bibfnamefont {R.}~\bibnamefont {Gupta}},
  \bibinfo {author} {\bibfnamefont {D.}~\bibnamefont {Das}}, \bibinfo {author}
  {\bibfnamefont {A.}~\bibnamefont {Leithe-Jasper}},\ and\ \bibinfo {author}
  {\bibfnamefont {E.}~\bibnamefont {Svanidze}},\ }\href@noop {} {\bibfield
  {journal} {\bibinfo  {journal} {Physical Review B}\ }\textbf {\bibinfo
  {volume} {102}},\ \bibinfo {pages} {014514} (\bibinfo {year}
  {2020}{\natexlab{b}})}\BibitemShut {NoStop}%
\bibitem [{\citenamefont {Tinkham}(2004)}]{tinkham2004introduction}%
  \BibitemOpen
  \bibfield  {author} {\bibinfo {author} {\bibfnamefont {M.}~\bibnamefont
  {Tinkham}},\ }\href@noop {} {\emph {\bibinfo {title} {Introduction to
  superconductivity}}}\ (\bibinfo  {publisher} {Courier Corporation},\ \bibinfo
  {year} {2004})\BibitemShut {NoStop}%
\bibitem [{\citenamefont {Annett}\ \emph {et~al.}(2004)\citenamefont {Annett}
  \emph {et~al.}}]{annett2004superconductivity}%
  \BibitemOpen
  \bibfield  {author} {\bibinfo {author} {\bibfnamefont {J.~F.}\ \bibnamefont
  {Annett}} \emph {et~al.},\ }\href@noop {} {\emph {\bibinfo {title}
  {Superconductivity, superfluids and condensates}}},\ Vol.~\bibinfo {volume}
  {5}\ (\bibinfo  {publisher} {Oxford University Press},\ \bibinfo {year}
  {2004})\BibitemShut {NoStop}%
\bibitem [{\citenamefont {Kr{\"a}geloh}(1969)}]{krageloh1969flux}%
  \BibitemOpen
  \bibfield  {author} {\bibinfo {author} {\bibfnamefont {U.}~\bibnamefont
  {Kr{\"a}geloh}},\ }\href@noop {} {\bibfield  {journal} {\bibinfo  {journal}
  {Physics Letters A}\ }\textbf {\bibinfo {volume} {28}},\ \bibinfo {pages}
  {657} (\bibinfo {year} {1969})}\BibitemShut {NoStop}%
\bibitem [{\citenamefont {Auer}\ and\ \citenamefont
  {Ullmaier}(1973)}]{auer1973magnetic}%
  \BibitemOpen
  \bibfield  {author} {\bibinfo {author} {\bibfnamefont {J.}~\bibnamefont
  {Auer}}\ and\ \bibinfo {author} {\bibfnamefont {H.}~\bibnamefont
  {Ullmaier}},\ }\href@noop {} {\bibfield  {journal} {\bibinfo  {journal}
  {Physical Review B}\ }\textbf {\bibinfo {volume} {7}},\ \bibinfo {pages}
  {136} (\bibinfo {year} {1973})}\BibitemShut {NoStop}%
\bibitem [{\citenamefont {Babaev}\ \emph {et~al.}(2017)\citenamefont {Babaev},
  \citenamefont {Carlstr{\"o}m}, \citenamefont {Silaev},\ and\ \citenamefont
  {Speight}}]{babaev2017type}%
  \BibitemOpen
  \bibfield  {author} {\bibinfo {author} {\bibfnamefont {E.}~\bibnamefont
  {Babaev}}, \bibinfo {author} {\bibfnamefont {J.}~\bibnamefont
  {Carlstr{\"o}m}}, \bibinfo {author} {\bibfnamefont {M.}~\bibnamefont
  {Silaev}},\ and\ \bibinfo {author} {\bibfnamefont {J.}~\bibnamefont
  {Speight}},\ }\href@noop {} {\bibfield  {journal} {\bibinfo  {journal}
  {Physica C: Superconductivity and its Applications}\ }\textbf {\bibinfo
  {volume} {533}},\ \bibinfo {pages} {20} (\bibinfo {year} {2017})}\BibitemShut
  {NoStop}%
\bibitem [{\citenamefont {Samoilenka}\ and\ \citenamefont
  {Babaev}(2020)}]{samoilenka2020spiral}%
  \BibitemOpen
  \bibfield  {author} {\bibinfo {author} {\bibfnamefont {A.}~\bibnamefont
  {Samoilenka}}\ and\ \bibinfo {author} {\bibfnamefont {E.}~\bibnamefont
  {Babaev}},\ }\href@noop {} {\bibfield  {journal} {\bibinfo  {journal}
  {Physical Review B}\ }\textbf {\bibinfo {volume} {102}},\ \bibinfo {pages}
  {184517} (\bibinfo {year} {2020})}\BibitemShut {NoStop}%
\bibitem [{\citenamefont {Bardeen}\ \emph {et~al.}(1957)\citenamefont
  {Bardeen}, \citenamefont {Cooper},\ and\ \citenamefont
  {Schrieffer}}]{bardeen1957theory}%
  \BibitemOpen
  \bibfield  {author} {\bibinfo {author} {\bibfnamefont {J.}~\bibnamefont
  {Bardeen}}, \bibinfo {author} {\bibfnamefont {L.~N.}\ \bibnamefont
  {Cooper}},\ and\ \bibinfo {author} {\bibfnamefont {J.~R.}\ \bibnamefont
  {Schrieffer}},\ }\href@noop {} {\bibfield  {journal} {\bibinfo  {journal}
  {Physical review}\ }\textbf {\bibinfo {volume} {108}},\ \bibinfo {pages}
  {1175} (\bibinfo {year} {1957})}\BibitemShut {NoStop}%
\bibitem [{\citenamefont {Dynes}\ \emph {et~al.}(1978)\citenamefont {Dynes},
  \citenamefont {Narayanamurti},\ and\ \citenamefont
  {Garno}}]{dynes1978direct}%
  \BibitemOpen
  \bibfield  {author} {\bibinfo {author} {\bibfnamefont {R.}~\bibnamefont
  {Dynes}}, \bibinfo {author} {\bibfnamefont {V.}~\bibnamefont
  {Narayanamurti}},\ and\ \bibinfo {author} {\bibfnamefont {J.~P.}\
  \bibnamefont {Garno}},\ }\href@noop {} {\bibfield  {journal} {\bibinfo
  {journal} {Physical Review Letters}\ }\textbf {\bibinfo {volume} {41}},\
  \bibinfo {pages} {1509} (\bibinfo {year} {1978})}\BibitemShut {NoStop}%
\bibitem [{\citenamefont {Tanaka}\ and\ \citenamefont
  {Kashiwaya}(1995)}]{tanaka1995theory}%
  \BibitemOpen
  \bibfield  {author} {\bibinfo {author} {\bibfnamefont {Y.}~\bibnamefont
  {Tanaka}}\ and\ \bibinfo {author} {\bibfnamefont {S.}~\bibnamefont
  {Kashiwaya}},\ }\href@noop {} {\bibfield  {journal} {\bibinfo  {journal}
  {Physical review letters}\ }\textbf {\bibinfo {volume} {74}},\ \bibinfo
  {pages} {3451} (\bibinfo {year} {1995})}\BibitemShut {NoStop}%
\bibitem [{\citenamefont {Suhl}\ \emph {et~al.}(1959)\citenamefont {Suhl},
  \citenamefont {Matthias},\ and\ \citenamefont {Walker}}]{suhl1959bardeen}%
  \BibitemOpen
  \bibfield  {author} {\bibinfo {author} {\bibfnamefont {H.}~\bibnamefont
  {Suhl}}, \bibinfo {author} {\bibfnamefont {B.}~\bibnamefont {Matthias}},\
  and\ \bibinfo {author} {\bibfnamefont {L.}~\bibnamefont {Walker}},\
  }\href@noop {} {\bibfield  {journal} {\bibinfo  {journal} {Physical Review
  Letters}\ }\textbf {\bibinfo {volume} {3}},\ \bibinfo {pages} {552} (\bibinfo
  {year} {1959})}\BibitemShut {NoStop}%
\bibitem [{\citenamefont {Szab{\'o}}\ \emph {et~al.}(2001)\citenamefont
  {Szab{\'o}}, \citenamefont {Samuely}, \citenamefont
  {Ka{\v{c}}mar{\v{c}}{\'\i}k}, \citenamefont {Klein}, \citenamefont {Marcus},
  \citenamefont {Fruchart}, \citenamefont {Miraglia}, \citenamefont
  {Marcenat},\ and\ \citenamefont {Jansen}}]{szabo2001evidence}%
  \BibitemOpen
  \bibfield  {author} {\bibinfo {author} {\bibfnamefont {P.}~\bibnamefont
  {Szab{\'o}}}, \bibinfo {author} {\bibfnamefont {P.}~\bibnamefont {Samuely}},
  \bibinfo {author} {\bibfnamefont {J.}~\bibnamefont
  {Ka{\v{c}}mar{\v{c}}{\'\i}k}}, \bibinfo {author} {\bibfnamefont
  {T.}~\bibnamefont {Klein}}, \bibinfo {author} {\bibfnamefont
  {J.}~\bibnamefont {Marcus}}, \bibinfo {author} {\bibfnamefont
  {D.}~\bibnamefont {Fruchart}}, \bibinfo {author} {\bibfnamefont
  {S.}~\bibnamefont {Miraglia}}, \bibinfo {author} {\bibfnamefont
  {C.}~\bibnamefont {Marcenat}},\ and\ \bibinfo {author} {\bibfnamefont
  {A.}~\bibnamefont {Jansen}},\ }\href@noop {} {\bibfield  {journal} {\bibinfo
  {journal} {Physical review letters}\ }\textbf {\bibinfo {volume} {87}},\
  \bibinfo {pages} {137005} (\bibinfo {year} {2001})}\BibitemShut {NoStop}%
\bibitem [{\citenamefont {Giubileo}\ \emph {et~al.}(2001)\citenamefont
  {Giubileo}, \citenamefont {Roditchev}, \citenamefont {Sacks}, \citenamefont
  {Lamy}, \citenamefont {Thanh}, \citenamefont {Klein}, \citenamefont
  {Miraglia}, \citenamefont {Fruchart}, \citenamefont {Marcus},\ and\
  \citenamefont {Monod}}]{giubileo2001two}%
  \BibitemOpen
  \bibfield  {author} {\bibinfo {author} {\bibfnamefont {F.}~\bibnamefont
  {Giubileo}}, \bibinfo {author} {\bibfnamefont {D.}~\bibnamefont {Roditchev}},
  \bibinfo {author} {\bibfnamefont {W.}~\bibnamefont {Sacks}}, \bibinfo
  {author} {\bibfnamefont {R.}~\bibnamefont {Lamy}}, \bibinfo {author}
  {\bibfnamefont {D.}~\bibnamefont {Thanh}}, \bibinfo {author} {\bibfnamefont
  {J.}~\bibnamefont {Klein}}, \bibinfo {author} {\bibfnamefont
  {S.}~\bibnamefont {Miraglia}}, \bibinfo {author} {\bibfnamefont
  {D.}~\bibnamefont {Fruchart}}, \bibinfo {author} {\bibfnamefont
  {J.}~\bibnamefont {Marcus}},\ and\ \bibinfo {author} {\bibfnamefont
  {P.}~\bibnamefont {Monod}},\ }\href@noop {} {\bibfield  {journal} {\bibinfo
  {journal} {Physical review letters}\ }\textbf {\bibinfo {volume} {87}},\
  \bibinfo {pages} {177008} (\bibinfo {year} {2001})}\BibitemShut {NoStop}%
\bibitem [{\citenamefont {Schmidt}\ \emph {et~al.}(2002)\citenamefont
  {Schmidt}, \citenamefont {Zasadzinski}, \citenamefont {Gray},\ and\
  \citenamefont {Hinks}}]{schmidt2002evidence}%
  \BibitemOpen
  \bibfield  {author} {\bibinfo {author} {\bibfnamefont {H.}~\bibnamefont
  {Schmidt}}, \bibinfo {author} {\bibfnamefont {J.}~\bibnamefont
  {Zasadzinski}}, \bibinfo {author} {\bibfnamefont {K.}~\bibnamefont {Gray}},\
  and\ \bibinfo {author} {\bibfnamefont {D.}~\bibnamefont {Hinks}},\
  }\href@noop {} {\bibfield  {journal} {\bibinfo  {journal} {Physical review
  letters}\ }\textbf {\bibinfo {volume} {88}},\ \bibinfo {pages} {127002}
  (\bibinfo {year} {2002})}\BibitemShut {NoStop}%
\bibitem [{\citenamefont {Iavarone}\ \emph {et~al.}(2002)\citenamefont
  {Iavarone}, \citenamefont {Karapetrov}, \citenamefont {Koshelev},
  \citenamefont {Kwok}, \citenamefont {Crabtree}, \citenamefont {Hinks},
  \citenamefont {Kang}, \citenamefont {Choi}, \citenamefont {Kim},
  \citenamefont {Kim} \emph {et~al.}}]{iavarone2002two}%
  \BibitemOpen
  \bibfield  {author} {\bibinfo {author} {\bibfnamefont {M.}~\bibnamefont
  {Iavarone}}, \bibinfo {author} {\bibfnamefont {G.}~\bibnamefont
  {Karapetrov}}, \bibinfo {author} {\bibfnamefont {A.}~\bibnamefont
  {Koshelev}}, \bibinfo {author} {\bibfnamefont {W.}~\bibnamefont {Kwok}},
  \bibinfo {author} {\bibfnamefont {G.}~\bibnamefont {Crabtree}}, \bibinfo
  {author} {\bibfnamefont {D.}~\bibnamefont {Hinks}}, \bibinfo {author}
  {\bibfnamefont {W.}~\bibnamefont {Kang}}, \bibinfo {author} {\bibfnamefont
  {E.-M.}\ \bibnamefont {Choi}}, \bibinfo {author} {\bibfnamefont {H.~J.}\
  \bibnamefont {Kim}}, \bibinfo {author} {\bibfnamefont {H.-J.}\ \bibnamefont
  {Kim}}, \emph {et~al.},\ }\href@noop {} {\bibfield  {journal} {\bibinfo
  {journal} {Physical review letters}\ }\textbf {\bibinfo {volume} {89}},\
  \bibinfo {pages} {187002} (\bibinfo {year} {2002})}\BibitemShut {NoStop}%
\bibitem [{\citenamefont {Gonnelli}\ \emph {et~al.}(2002)\citenamefont
  {Gonnelli}, \citenamefont {Daghero}, \citenamefont {Ummarino}, \citenamefont
  {Stepanov}, \citenamefont {Jun}, \citenamefont {Kazakov},\ and\ \citenamefont
  {Karpinski}}]{gonnelli2002direct}%
  \BibitemOpen
  \bibfield  {author} {\bibinfo {author} {\bibfnamefont {R.}~\bibnamefont
  {Gonnelli}}, \bibinfo {author} {\bibfnamefont {D.}~\bibnamefont {Daghero}},
  \bibinfo {author} {\bibfnamefont {G.}~\bibnamefont {Ummarino}}, \bibinfo
  {author} {\bibfnamefont {V.}~\bibnamefont {Stepanov}}, \bibinfo {author}
  {\bibfnamefont {J.}~\bibnamefont {Jun}}, \bibinfo {author} {\bibfnamefont
  {S.}~\bibnamefont {Kazakov}},\ and\ \bibinfo {author} {\bibfnamefont
  {J.}~\bibnamefont {Karpinski}},\ }\href@noop {} {\bibfield  {journal}
  {\bibinfo  {journal} {Physical review letters}\ }\textbf {\bibinfo {volume}
  {89}},\ \bibinfo {pages} {247004} (\bibinfo {year} {2002})}\BibitemShut
  {NoStop}%
\bibitem [{\citenamefont {Silva-Guill{\'e}n}\ \emph {et~al.}(2015)\citenamefont
  {Silva-Guill{\'e}n}, \citenamefont {Noat}, \citenamefont {Cren},
  \citenamefont {Sacks}, \citenamefont {Canadell},\ and\ \citenamefont
  {Ordej{\'o}n}}]{silva2015tunneling}%
  \BibitemOpen
  \bibfield  {author} {\bibinfo {author} {\bibfnamefont {J.}~\bibnamefont
  {Silva-Guill{\'e}n}}, \bibinfo {author} {\bibfnamefont {Y.}~\bibnamefont
  {Noat}}, \bibinfo {author} {\bibfnamefont {T.}~\bibnamefont {Cren}}, \bibinfo
  {author} {\bibfnamefont {W.}~\bibnamefont {Sacks}}, \bibinfo {author}
  {\bibfnamefont {E.}~\bibnamefont {Canadell}},\ and\ \bibinfo {author}
  {\bibfnamefont {P.}~\bibnamefont {Ordej{\'o}n}},\ }\href@noop {} {\bibfield
  {journal} {\bibinfo  {journal} {Physical Review B}\ }\textbf {\bibinfo
  {volume} {92}},\ \bibinfo {pages} {064514} (\bibinfo {year}
  {2015})}\BibitemShut {NoStop}%
\bibitem [{\citenamefont {Schopohl}\ and\ \citenamefont
  {Scharnberg}(1977)}]{schopohl1977tunneling}%
  \BibitemOpen
  \bibfield  {author} {\bibinfo {author} {\bibfnamefont {N.}~\bibnamefont
  {Schopohl}}\ and\ \bibinfo {author} {\bibfnamefont {K.}~\bibnamefont
  {Scharnberg}},\ }\href@noop {} {\bibfield  {journal} {\bibinfo  {journal}
  {Solid State Communications}\ }\textbf {\bibinfo {volume} {22}},\ \bibinfo
  {pages} {371} (\bibinfo {year} {1977})}\BibitemShut {NoStop}%
\bibitem [{\citenamefont {Noat}\ \emph {et~al.}(2010)\citenamefont {Noat},
  \citenamefont {Cren}, \citenamefont {Debontridder}, \citenamefont
  {Roditchev}, \citenamefont {Sacks}, \citenamefont {Toulemonde},\ and\
  \citenamefont {San~Miguel}}]{noat2010signatures}%
  \BibitemOpen
  \bibfield  {author} {\bibinfo {author} {\bibfnamefont {Y.}~\bibnamefont
  {Noat}}, \bibinfo {author} {\bibfnamefont {T.}~\bibnamefont {Cren}}, \bibinfo
  {author} {\bibfnamefont {F.}~\bibnamefont {Debontridder}}, \bibinfo {author}
  {\bibfnamefont {D.}~\bibnamefont {Roditchev}}, \bibinfo {author}
  {\bibfnamefont {W.}~\bibnamefont {Sacks}}, \bibinfo {author} {\bibfnamefont
  {P.}~\bibnamefont {Toulemonde}},\ and\ \bibinfo {author} {\bibfnamefont
  {A.}~\bibnamefont {San~Miguel}},\ }\href@noop {} {\bibfield  {journal}
  {\bibinfo  {journal} {Physical Review B}\ }\textbf {\bibinfo {volume} {82}},\
  \bibinfo {pages} {014531} (\bibinfo {year} {2010})}\BibitemShut {NoStop}%
\bibitem [{\citenamefont {Eskildsen}\ \emph {et~al.}(2002)\citenamefont
  {Eskildsen}, \citenamefont {Kugler}, \citenamefont {Tanaka}, \citenamefont
  {Jun}, \citenamefont {Kazakov}, \citenamefont {Karpinski},\ and\
  \citenamefont {Fischer}}]{eskildsen2002vortex}%
  \BibitemOpen
  \bibfield  {author} {\bibinfo {author} {\bibfnamefont {M.}~\bibnamefont
  {Eskildsen}}, \bibinfo {author} {\bibfnamefont {M.}~\bibnamefont {Kugler}},
  \bibinfo {author} {\bibfnamefont {S.}~\bibnamefont {Tanaka}}, \bibinfo
  {author} {\bibfnamefont {J.}~\bibnamefont {Jun}}, \bibinfo {author}
  {\bibfnamefont {S.}~\bibnamefont {Kazakov}}, \bibinfo {author} {\bibfnamefont
  {J.}~\bibnamefont {Karpinski}},\ and\ \bibinfo {author} {\bibfnamefont
  {{\O}.}~\bibnamefont {Fischer}},\ }\href@noop {} {\bibfield  {journal}
  {\bibinfo  {journal} {Physical review letters}\ }\textbf {\bibinfo {volume}
  {89}},\ \bibinfo {pages} {187003} (\bibinfo {year} {2002})}\BibitemShut
  {NoStop}%
\bibitem [{\citenamefont {Kohn}\ and\ \citenamefont
  {Sham}(1965)}]{kohn1965self}%
  \BibitemOpen
  \bibfield  {author} {\bibinfo {author} {\bibfnamefont {W.}~\bibnamefont
  {Kohn}}\ and\ \bibinfo {author} {\bibfnamefont {L.~J.}\ \bibnamefont
  {Sham}},\ }\href@noop {} {\bibfield  {journal} {\bibinfo  {journal} {Physical
  review}\ }\textbf {\bibinfo {volume} {140}},\ \bibinfo {pages} {A1133}
  (\bibinfo {year} {1965})}\BibitemShut {NoStop}%
\bibitem [{\citenamefont {Giannozzi}\ \emph {et~al.}(2009)\citenamefont
  {Giannozzi}, \citenamefont {Baroni}, \citenamefont {Bonini}, \citenamefont
  {Calandra}, \citenamefont {Car}, \citenamefont {Cavazzoni}, \citenamefont
  {Ceresoli}, \citenamefont {Chiarotti}, \citenamefont {Cococcioni},
  \citenamefont {Dabo} \emph {et~al.}}]{giannozzi2009quantum}%
  \BibitemOpen
  \bibfield  {author} {\bibinfo {author} {\bibfnamefont {P.}~\bibnamefont
  {Giannozzi}}, \bibinfo {author} {\bibfnamefont {S.}~\bibnamefont {Baroni}},
  \bibinfo {author} {\bibfnamefont {N.}~\bibnamefont {Bonini}}, \bibinfo
  {author} {\bibfnamefont {M.}~\bibnamefont {Calandra}}, \bibinfo {author}
  {\bibfnamefont {R.}~\bibnamefont {Car}}, \bibinfo {author} {\bibfnamefont
  {C.}~\bibnamefont {Cavazzoni}}, \bibinfo {author} {\bibfnamefont
  {D.}~\bibnamefont {Ceresoli}}, \bibinfo {author} {\bibfnamefont {G.~L.}\
  \bibnamefont {Chiarotti}}, \bibinfo {author} {\bibfnamefont {M.}~\bibnamefont
  {Cococcioni}}, \bibinfo {author} {\bibfnamefont {I.}~\bibnamefont {Dabo}},
  \emph {et~al.},\ }\href@noop {} {\bibfield  {journal} {\bibinfo  {journal}
  {Journal of physics: Condensed matter}\ }\textbf {\bibinfo {volume} {21}},\
  \bibinfo {pages} {395502} (\bibinfo {year} {2009})}\BibitemShut {NoStop}%
\bibitem [{\citenamefont {Perdew}\ \emph {et~al.}(1996)\citenamefont {Perdew},
  \citenamefont {Burke},\ and\ \citenamefont
  {Ernzerhof}}]{perdew1996generalized}%
  \BibitemOpen
  \bibfield  {author} {\bibinfo {author} {\bibfnamefont {J.~P.}\ \bibnamefont
  {Perdew}}, \bibinfo {author} {\bibfnamefont {K.}~\bibnamefont {Burke}},\ and\
  \bibinfo {author} {\bibfnamefont {M.}~\bibnamefont {Ernzerhof}},\ }\href@noop
  {} {\bibfield  {journal} {\bibinfo  {journal} {Physical review letters}\
  }\textbf {\bibinfo {volume} {77}},\ \bibinfo {pages} {3865} (\bibinfo {year}
  {1996})}\BibitemShut {NoStop}%
\bibitem [{\citenamefont {Bl{\"o}chl}(1994)}]{blochl1994projector}%
  \BibitemOpen
  \bibfield  {author} {\bibinfo {author} {\bibfnamefont {P.~E.}\ \bibnamefont
  {Bl{\"o}chl}},\ }\href@noop {} {\bibfield  {journal} {\bibinfo  {journal}
  {Physical review B}\ }\textbf {\bibinfo {volume} {50}},\ \bibinfo {pages}
  {17953} (\bibinfo {year} {1994})}\BibitemShut {NoStop}%
\bibitem [{\citenamefont {Monkhorst}\ and\ \citenamefont
  {Pack}(1976)}]{monkhorst1976special}%
  \BibitemOpen
  \bibfield  {author} {\bibinfo {author} {\bibfnamefont {H.~J.}\ \bibnamefont
  {Monkhorst}}\ and\ \bibinfo {author} {\bibfnamefont {J.~D.}\ \bibnamefont
  {Pack}},\ }\href@noop {} {\bibfield  {journal} {\bibinfo  {journal} {Physical
  review B}\ }\textbf {\bibinfo {volume} {13}},\ \bibinfo {pages} {5188}
  (\bibinfo {year} {1976})}\BibitemShut {NoStop}%
\bibitem [{\citenamefont {Koshelev}\ and\ \citenamefont
  {Golubov}(2003)}]{koshelev2003mixed}%
  \BibitemOpen
  \bibfield  {author} {\bibinfo {author} {\bibfnamefont {A.}~\bibnamefont
  {Koshelev}}\ and\ \bibinfo {author} {\bibfnamefont {A.~A.}\ \bibnamefont
  {Golubov}},\ }\href@noop {} {\bibfield  {journal} {\bibinfo  {journal}
  {Physical review letters}\ }\textbf {\bibinfo {volume} {90}},\ \bibinfo
  {pages} {177002} (\bibinfo {year} {2003})}\BibitemShut {NoStop}%
\bibitem [{\citenamefont {Silaev}\ and\ \citenamefont
  {Babaev}(2011)}]{silaev2011microscopic}%
  \BibitemOpen
  \bibfield  {author} {\bibinfo {author} {\bibfnamefont {M.}~\bibnamefont
  {Silaev}}\ and\ \bibinfo {author} {\bibfnamefont {E.}~\bibnamefont
  {Babaev}},\ }\href@noop {} {\bibfield  {journal} {\bibinfo  {journal}
  {Physical Review B}\ }\textbf {\bibinfo {volume} {84}},\ \bibinfo {pages}
  {094515} (\bibinfo {year} {2011})}\BibitemShut {NoStop}%
\bibitem [{\citenamefont {Vagov}\ \emph {et~al.}(2016)\citenamefont {Vagov},
  \citenamefont {Shanenko}, \citenamefont {Milo{\v{s}}evi{\'c}}, \citenamefont
  {Axt}, \citenamefont {Vinokur}, \citenamefont {Aguiar},\ and\ \citenamefont
  {Peeters}}]{vagov2016superconductivity}%
  \BibitemOpen
  \bibfield  {author} {\bibinfo {author} {\bibfnamefont {A.}~\bibnamefont
  {Vagov}}, \bibinfo {author} {\bibfnamefont {A.}~\bibnamefont {Shanenko}},
  \bibinfo {author} {\bibfnamefont {M.}~\bibnamefont {Milo{\v{s}}evi{\'c}}},
  \bibinfo {author} {\bibfnamefont {V.~M.}\ \bibnamefont {Axt}}, \bibinfo
  {author} {\bibfnamefont {V.}~\bibnamefont {Vinokur}}, \bibinfo {author}
  {\bibfnamefont {J.~A.}\ \bibnamefont {Aguiar}},\ and\ \bibinfo {author}
  {\bibfnamefont {F.}~\bibnamefont {Peeters}},\ }\href@noop {} {\bibfield
  {journal} {\bibinfo  {journal} {Physical Review B}\ }\textbf {\bibinfo
  {volume} {93}},\ \bibinfo {pages} {174503} (\bibinfo {year}
  {2016})}\BibitemShut {NoStop}%
\end{thebibliography}%

\begin{figure}[h!]
	\centering
	\includegraphics[width=0.75\textwidth]{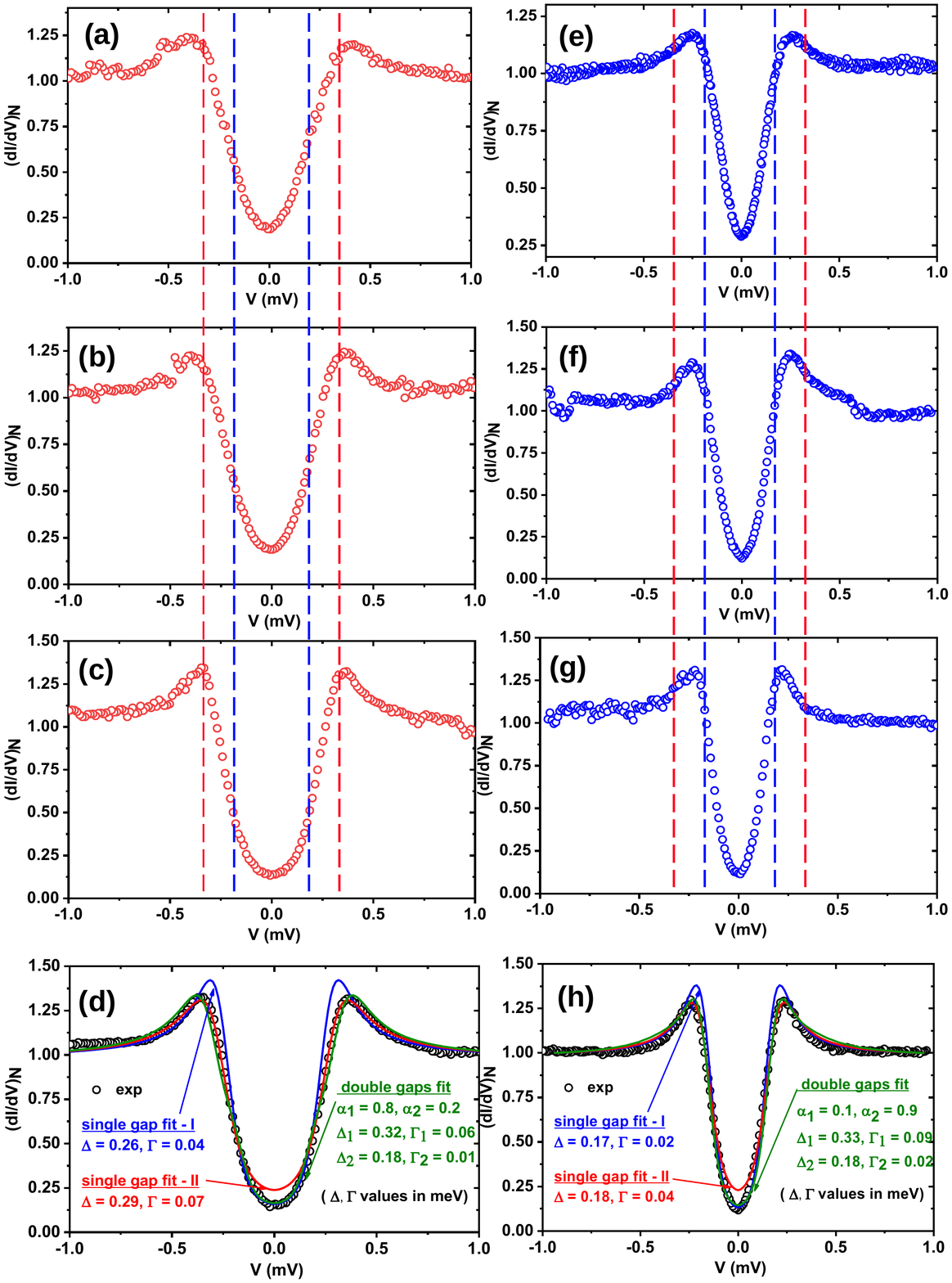}
	\caption{ Eight representative tunneling spectra probed at different points on the sample surface, all recorded at $T$ $\sim$ 310 mK. The spectra (red circles) represented in \textbf{(a)-(d)} have the dominant contribution from the larger gap ($\Delta_1$ $\simeq$ 320$\pm$10 $\mu$eV), and those represented in \textbf{(e)-(h)} have the same from the smaller gap ($\Delta_2$ $\simeq$ 180$\pm$10 $\mu$eV). The verticle lines are guide to the eye and correspond to the biasing of $\pm$320 $\mu$V (red) and $\pm$180 $\mu$V (blue), respectively. For visual clarity, best fittings within a single-gap $s$-wave model (red lines matching the upper part only, and the blue lines matching the lower part only) and the two-gap $s$-wave model (green lines matching the whole spectrum) are presented in (d) and (h) only.}
	
\end{figure}

\begin{figure}[h!]
	\centering
	\includegraphics[width=0.7\textwidth]{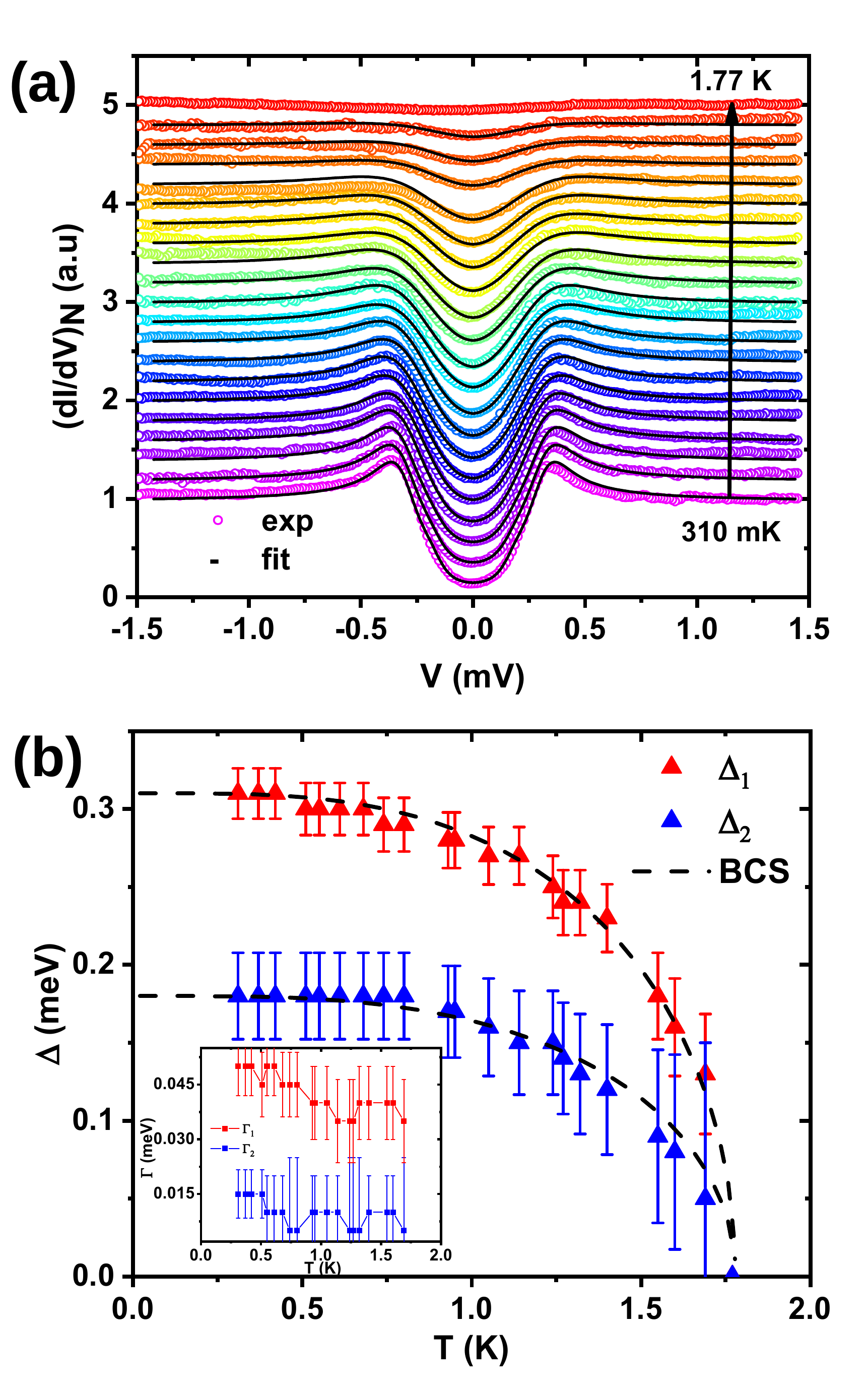}
	\caption{\textbf{(a)} The temperature dependence of a typical tunneling conductance spectra with the theoretical fits in the absence of any magnetic field. At the lowest $T$, $\Delta_1$=310 $\mu$eV, $\Delta_2$=180 $\mu$eV, $\Gamma_1$=50 $\mu$eV, $\Gamma_2$=15 $\mu$eV for the spectrum. The relative contributions $\alpha_1$=0.8 and $\alpha_2$=0.2 remain constant throughout the $T$ range. \textbf{(b)} Evolution of the two gaps ($\Delta_1$ and $\Delta_2$) with $T$, extracted from the plot (a) along with two-gaps BCS fits. In the inset: the $T$ evolution of the two broadening parameters ($\Gamma_1$ and $\Gamma_2$) associated with the two gaps extracted similarly. The errors of the fitting parameters are shown in the respective plots.}
	
\end{figure}

\begin{figure}[h!]
	\centering
	\includegraphics[width=0.7\textwidth]{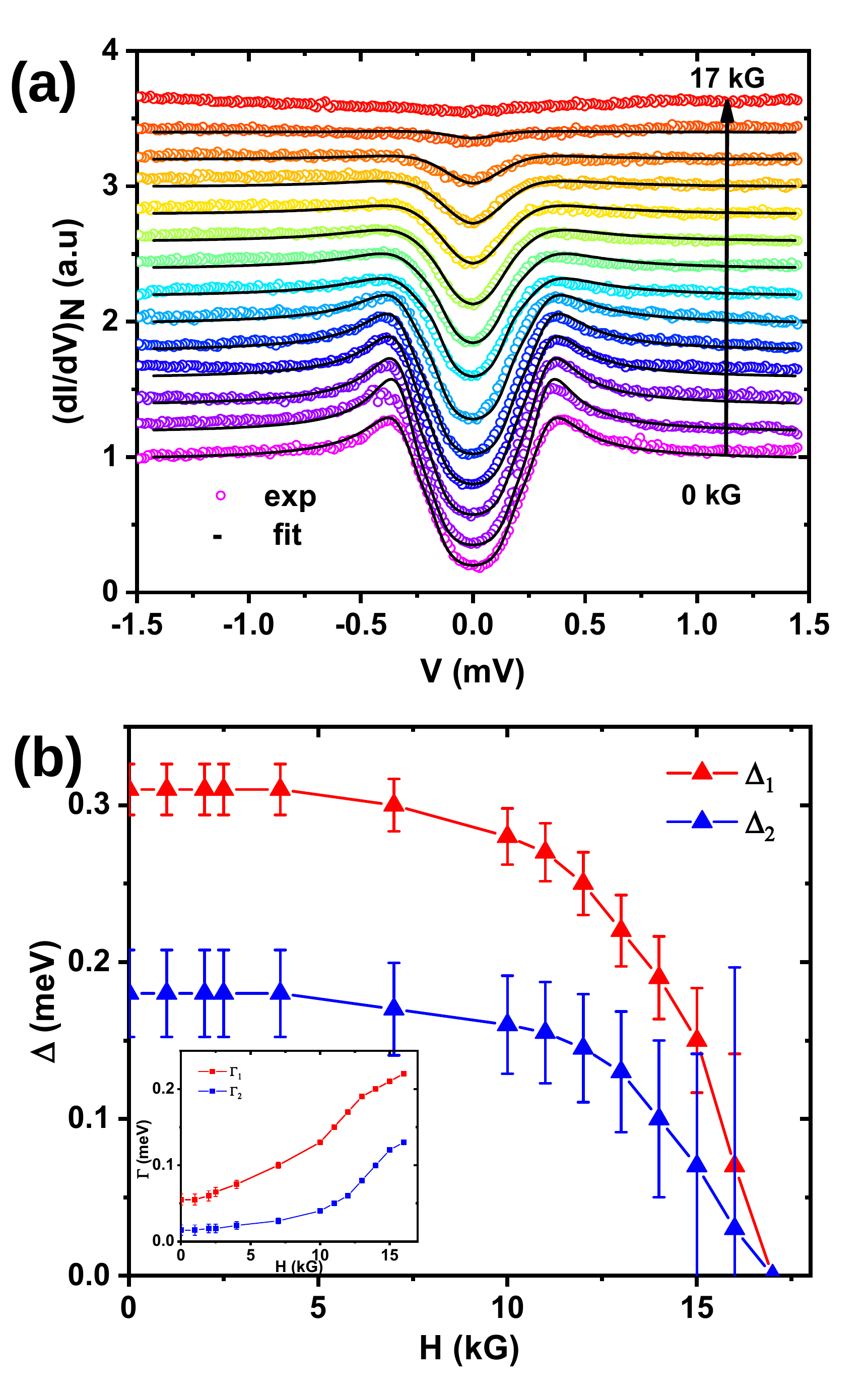}
	\caption{\textbf{(a)} Magnetic field dependence of tunneling conductance spectra recorded and kept throughout at $T$ = 310 mK along with corresponding theoretical fits. \textbf{(b)} Evolution of $\Delta_1$ and $\Delta_2$ with the magnetic field extracted from the plot (a). $Inset$: the field evolution of the two broadening parameters ($\Gamma_1$ and $\Gamma_2$) associated with the two gaps. The fitting errors of the parameters are also shown in the respective plots. The same spectrum is used for $T$ and $H$ dependence keeping the position of the tip unchanged on the sample surface.}
	
\end{figure}

\begin{figure}[h!]
	\centering
	\includegraphics[width=0.68\textwidth]{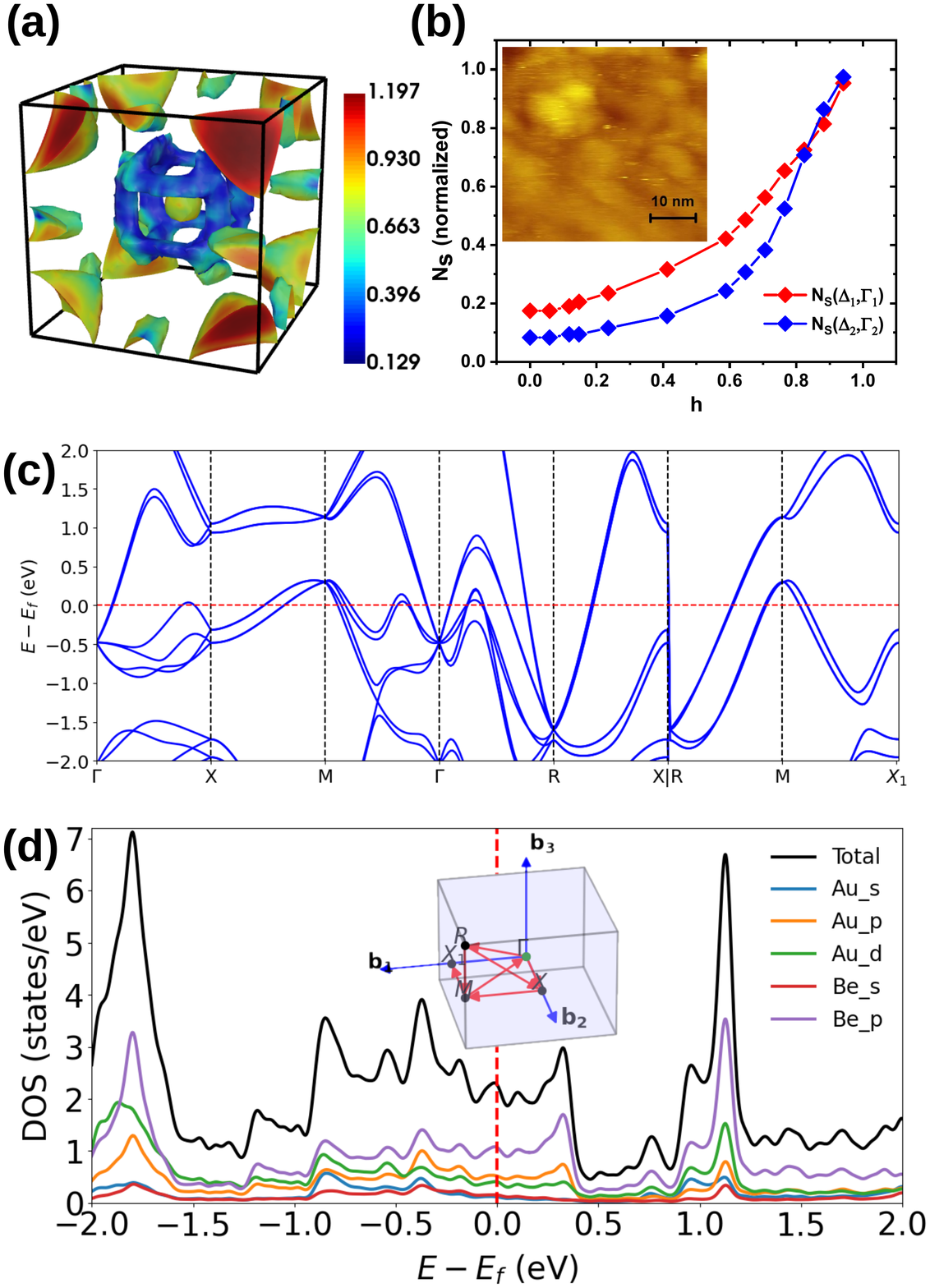}
	\caption{\textbf{(a)} The Fermi surface of AuBe with the Fermi velocities at different pockets represented with color gradients. \textbf{(b)} Normalized zero-bias DOS corresponding to two different bands as a function of the reduced magnetic field $h$ = $H$/$H_{c(l)}$. $H_{c(l)}$ = 17 kG is the spectroscopically measured local critical field. The normalization is done by dividing the $[dI/dV]_{V=0}$ $_{meV}$ with $[dI/dV]_{V=1.5}$ $_{meV}$. $Inset$: An STM topographic image of the sample surface. \textbf{(c)} The band dispersion along the high symmetry directions, where band split is visible due to moderate spin-orbit coupling. For visual clarity, a slight doubling of the surfaces due to this band splitting is not presented in (a). \textbf{(d)} total and orbital-resolved DOS considering the spin-orbit coupling. $Inset$: the high symmetry paths joining the high symmetry points within the first Brillouin zone of simple-cubic shape.}
	
\end{figure}

\end{document}